**Efficiency limits in in-line imaging**

T. E. Gureyev [a)], D. M. Paganin [b)] and H. M. Quiney [a)]

[a)] School of Physics, University of Melbourne, Parkville, Victoria, 3010, Australia

[b)] School of Physics and Astronomy, Monash University, Clayton, Victoria, 3800, Australia

Correspondence email: timur.gureyev@unimelb.edu.au

## Abstract

In-line phase-contrast imaging, also known as propagation-based imaging or Gabor holography, is capable of delivering significant gains in image quality, compared to attenuation-based imaging at the same radiation dose. Image quality metrics, including signal-to-noise ratio and spatial resolution, are considered, and their relationship to the amount of Shannon information in the image is discussed. It is demonstrated that it is possible to increase the image quality gain by combining known phase retrieval methods with deconvolution of the point-spread function (PSF) of the imaging system, with the highest gain achieved using the Contrast Transfer Function correction method in combination with the PSF deconvolution. The ultimate limit of this gain is determined by the ratio of the real decrement to the imaginary part of the complex refractive index of the imaged sample. Analytical results for the image quality limits in in-line imaging are obtained using a Gaussian model for the detector PSF. A numerically simulated example confirming the theoretical results is also presented.

## 1. Introduction

Phase-contrast imaging methods, utilizing coherent free-space propagation of radiation or matter waves after their interaction with an imaged object, are capable of delivering superior contrast-to-noise ratio (CNR) compared to attenuation-based imaging (ABI) at the same radiation dose and spatial resolution. A class of such imaging techniques that do not involve any optical elements between the imaged object and detector are known as in-line or propagation-based imaging (PBI). These techniques are closely related to Gabor holography (Gabor, 1948) and defocus-series imaging (Schiske, 1968, 2002). Imaging methods of this general type are used in electron microscopy (Saxton, 1994), including cryo-EM (Glaeser, 2019), optical imaging (Bremmer, 1952; Teague, 1983), X-ray imaging (Paganin, 2006; Wilkins et al., 2014), neutron radiography (Allman et al., 2000) and other fields.

In recent years, PBI techniques have become particularly popular in macroscopic and microscopic X-ray imaging (Snigirev et al., 1995; Wilkins et al., 1996; Nugent et al., 1996; Cloetens et al., 1996; Endrizzi, 2018). In the present paper, we primarily target X-ray PBI of biological samples, while noting that our methods and results are much more broadly applicable. X-ray PBI is currently being adapted for medical imaging of excised tissue specimens, as well as for screening and diagnostic imaging of live patients, particularly for breast and lung cancer (Bravin et al., 2013; Longo et al., 2019; Momose, 2020; Arhatari et al., 2021). The volumetric three-dimensional (3D) implementation of PBI is usually referred to as propagation-based computed tomography (PB-CT). Compared to other X-ray phase-contrast imaging methods, PBI is simpler in principle as it does not require any extra optical elements in addition to an X-ray source, sample stage and a 2D position-sensitive detector[1]. However, PBI requires the X-ray beam incident on the sample to be sufficiently spatially coherent in order to be able to detect the phase contrast (Wilkins et al., 1996). This typically restricts the use of the technique to synchrotron beamlines or laboratory microfocus X-ray sources. This limitation has obvious practical implications, such as the high cost and low availability of synchrotron facilities on one hand, and insufficient brightness of most commercially available microfocus X-ray sources for enabling medical imaging with acceptably short exposure times. Nevertheless, projects are currently in progress involving construction of specialized medical imaging facilities utilizing PBI at synchrotrons (Longo et al., 2019; Arhatari et al., 2021). Technology is also forging ahead in producing more powerful laboratory X-ray sources with effective source sizes of the order of 10 μm, as typically required for biomedical applications of PBI (Espes et al., 2014; Gambari et al., 2020; Van Elk, 2025). On the other hand, the use of artificial intelligence (AI) in X-ray image processing and analysis is showing promise in delivering high diagnostic imaging quality at significantly reduced levels of incident X-ray fluence (radiation dose) (Chen et al., 2017; Pakzad et al., 2026; Tajbakhsh et al., 2026). This may eventually enable PBI imaging with acceptable exposure times for medical purposes using less powerful microfocus X-ray sources.

[1] This method of X-ray phase contrast imaging may be viewed as a form of inline holography in which the sample-to-detector distance is sufficiently small to be in the (near) Fresnel regime. The method exhibits phase contrast in the sense that the propagated intensity image is a function of the phase of the pre-propagation wave field over the nominally planar exit surface of the sample. For hard X-rays illuminating amorphous samples whose size is on the order of (say) a few millimetres, typical sample-to-detector distances are on the order of one metre. In such near-Fresnel-regime images, there is typically just one dominant Fresnel fringe, with the diffraction-induced transverse energy flow being on the order of a few pixels or less. This implies the morphology of the measured phase-contrast image to be closely related to the morphology of the sample when projected along the direction of the optical axis.

Development of medical applications of PBI stands to benefit from optimization of relevant imaging setups and image processing methods. In modern practice, the line between the hardware components used for image acquisition and the software for image processing is often blurred, and it becomes necessary to study and optimize the computational imaging system as a whole (Paganin et al., 2004b; Mait et al., 2018; Gureyev et al., 2019). In the context of PBI and PB-CT, the two key tasks are: (1) optimization of the imaging hardware setup, including the X-ray source, the detector and the geometrical parameters of the setup (Gureyev et al., 2026a), and (2) optimization of the image processing algorithms, including phase retrieval and CT reconstruction (Gureyev et al., 2026b). These two tasks are inter-dependent, in the sense that the optimization of image acquisition setup depends on the image processing strategy, and vice versa. The present paper is focused on the phase retrieval step of this computational imaging pipeline. In the area of biomedical applications of PBI and PB-CT, the most popular approach to phase retrieval is currently Paganin's method, which is based on the homogeneous Transport of Intensity equation (TIE-Hom) (Paganin et al., 2002; Paganin, 2006)[2]. The TIE-Hom approach is capable of delivering images with significantly higher CNR compared to ABI at the same radiation dose and spatial resolution (Nesterets et al., 2014). The corresponding gain in imaging quality[3] can be defined as the ratio of SNR in

[2] (i) The transport-of-intensity equation (TIE) is a continuity equation for paraxial monochromatic scalar light beams (Teague, 1983). It expresses the rate of change of intensity with respect to propagation distance along the optical axis, in terms of the intensity and phase of the beam over planes perpendicular to that axis. (ii) The TIE-Hom is a special case wherein a single-material sample is considered to be coherently illuminated by planar or spherical waves. For example, when imaging biological soft tissue using hard X-rays, the fact that such high-energy photons interact primarily with atomic electrons means that — for the purpose of beam-sample interactions — the sample may often be approximated by a single variable-density material, namely electrons. Note, moreover, that the term "homogeneous" in "TIE-Hom" actually represents a generalization of the "single material" concept. We shall further explain this latter point in due course. (iii) The single-material assumption in TIE-Hom couples the phase and intensity of the X-ray beam over the nominally planar exit surface of the illuminated sample, wherein the exit-surface intensity is a simple function of the exit-surface phase, and conversely. This simplifies the required data collection by halving the number of required intensity measurements. This simplicity is one of the reasons that the method has been widely applied in coherent X-ray imaging, with another important reason being the extraordinary stability of the method with respect to noise in the input intensity measurements.

[3] (i) The "noisy-channel coding theorem" (Shannon 1948; MacKay, 2003) implies that, when a signal is transmitted through a noisy information channel, the SNR of the transmitted signal may be increased by suitably coding the signal before transmission and then decoding after transmission. (ii) As explained in Gureyev *et al.* (2024), the TIE-Hom approach is a special case of this strategy. In particular, refractive information regarding a

the images obtained using the TIE-Hom and the ABI methods, at the same spatial resolution and incident photon flux (Nesterets et al., 2014). This gain has been quantified (Nesterets et al., 2014; Kitchen et al., 2017; Gureyev et al., 2025, 2026a) and shown to be ultimately proportional to the ratio ($\gamma$) of the real decrement ($\delta$) to the imaginary part ($\beta$) of the complex refractive index of the imaged sample, $n = 1-\delta+i\beta$, $\gamma \equiv \delta/\beta$. This result can be understood in terms of the underlying imaging physics. Indeed, image quality, corresponding to the amount of Shannon information about the imaged object available in the image, is fundamentally determined by the interaction of X-ray photons with the imaged object (Gureyev et al., 2016). Image quality increases with the photon fluence, in proportion to the number of photons per unit projected area (in PBI) or unit volume (in PB-CT) of the sample. This dependence of the imaging quality on the X-ray fluence (or the radiation dose) is similar in attenuation and phase-contrast imaging. However, the amount of information about the object accessible in the resultant image also depends on the strength of interaction of incident X-ray photons with electrons in the sample. The relative interaction strength of the X-ray absorption and refraction channels can be expressed as a ratio of the corresponding scattering cross-sections, which is equal to $\gamma$, making this parameter a key determinant of imaging quality in PBI.

Let us consider more specifically the "intrinsic imaging quality metric" (Gureyev et al., 2014, 2016) that is equal (in the 2D case) to the ratio of SNR to the spatial resolution, divided by the square root of the incident photon fluence[4]. It has been shown that this metric is closely

---

sample that is imprinted upon the phase of an X-ray beam passing through it is encoded in the measured noisy phase-contrast image. Subsequent phase retrieval constitutes "decoding" in this context. Since such imaging is an example of a Shannon-type code-transmit-decode strategy, the logical possibility exists that a gain in SNR may be operative, relative to the coding-free strategy which corresponds to measuring a contact image to determine the intensity (and thereby the phase) imprinted upon an X-ray beam traversing a homogeneous object. This is the "gain in imaging quality" referred to in the main text of the paper.

[4] To supplement the formal development in the cited papers, for the convenience of the reader we here recapitulate a heuristic argument motivating why the intrinsic imaging quality metric $Q$ has the indicated form. Let $\overline{I_{in}}$ be the fluence, namely the mean number of photons per unit area that would be detected were a planar-surface perfect detector to be employed. If such a detector were to have square pixels with physical width $\Delta_{det}$, then the mean number of photons $N$ that would be detected by such an idealized pixel would be the fluence multiplied by the pixel area, thus $N = \overline{I_{in}}(\Delta_{pix})^2$. Applying Poisson statistics to our perfect detector pixel implies that the corresponding idealized SNR is $SNR_{ideal} = N/\sqrt{N} = \sqrt{N} = \sqrt{\overline{I_{in}}}\Delta_{pix}$. Since the actual SNR will be no better than the idealized SNR, it is natural to define $Q$ as the ratio the actual SNR and the idealized SNR, thus $Q \equiv (SNR/SNR_{ideal}) \le 1$. Using the preceding expression for $SNR_{ideal}$ then gives $Q = SNR/(\sqrt{\overline{I_{in}}}\Delta_{pix})$, in accord with the sentence to which this footnote is attached.

related to the amount of detectable Shannon information per incident photon (Gureyev et al., 2016). It has also been shown (Gureyev et al., 2017, 2025) that, in the case of the 2D TIE-Hom phase retrieval method, the gain in the intrinsic imaging quality is proportional to $(\gamma / N_F)^{1/2}$, where $N_F = \Delta_{det}^2 / (R\lambda)$ is the Fresnel number corresponding to the spatial resolution $\Delta_{det}$ of the detector, $R$ is the X-ray propagation ("defocus") distance and $\lambda$ is the X-ray wavelength (considering monochromatic X-rays for simplicity). As $N_F$ must be larger than unity in order for the TIE-Hom approximation to be valid, the ratio $\gamma / N_F$ is always smaller than $\gamma$ in this method, indicating that the image quality gain may be sub-optimal. More generally, this result shows that the image quality gain in TIE-Hom depends on the parameters of imaging setup, rather than just on the scattering cross-sections. On the other hand, it was shown by means of numerical analysis in (Nesterets et al., 2014) that the gain in SNR in PBI can be as high as $0.3\gamma$ when a homogeneous version (CTF-Hom) of the Contrast Transfer Function formalism (Cowley, 1995; Pogany et al., 1997; Gureyev et al., 2014) is used instead of the TIE-Hom. This result is further investigated in the present paper.

Another question relevant to the present study is about the exact mechanism behind the "unreasonable" effectiveness of the TIE-Hom imaging that underpins its advantage in comparison with conventional absorption-based X-ray imaging methods (Gureyev et al., 2017). Details of this subject are rather technical and are studied in subsequent sections of the present paper. The key outcome of this investigation demonstrated below is that the explanation originally suggested in (Gureyev et al., 2017) and several later publications for the effectiveness of the TIE-Hom imaging was not entirely correct. It was proposed in (Gureyev et al., 2017) that the image quality improvement in PBI takes place at the stage of forward free-space propagation, and the quality then remains constant at the retrieval stage. More specifically, it was stated that the spatial resolution was improved at the forward propagation stage, while SNR was identical in the object and image planes due to the photon number conservation in free-space propagation. The subsequent TIE-Hom retrieval procedure returned the spatial resolution to the same level as in the object plane, but increased the SNR by the factor $(\gamma / N_F)^{1/2}$, leaving the ratio of the SNR to spatial resolution at the same level as in the image plane. This gave the overall gain in the image quality equal to $(\gamma / N_F)^{1/2}$, in comparison with ABI at the same incident photon fluence, after the complete process of forward free-space propagation, image detection in the "propagated" plane and TIE-Hom retrieval. It turns out that a weak link in this logical chain in (Gureyev et al., 2017) was the fact that the spatial resolution in the image plane was not explicitly calculated in the same form that was later used in the evaluation of the overall gain in the image quality. In the present paper, we use a simple Gaussian model for the detector point-spread function (PSF), which allows us to calculate the spatial resolution in the image plane explicitly. We then show that spatial resolution increases after a coherent free-space propagation by a factor proportional to $\gamma / N_F$, rather than $(\gamma / N_F)^{1/2}$ as claimed in (Gureyev et al., 2017). At the subsequent TIE-Hom retrieval step, the spatial resolution returns to the same level as in the object plane (i.e. reduces by the factor $\gamma / N_F$ compared to the image plane), but the SNR increases by $(\gamma / N_F)^{1/2}$. The overall gain in image quality, expressed via the ratio of SNR to spatial resolution, is thus equal to $(\gamma / N_F)^{1/2}$ in the complete PBI process with TIE-Hom retrieval, as stated in the previous publications. However, the intermediate gain in the image plane is actually larger, being equal to $\gamma / N_F$. This means that the TIE-Hom retrieval step actually lowers the image quality compared to the "raw" image registered in the "propagated" image plane, rather than keeping it constant as suggested in (Gureyev et al., 2017) and elsewhere.

The fact that the TIE-Hom retrieval lowers the image quality gain leads to the natural question about possible improvements to the "phase" retrieval stage of PBI and PB-CT, that could potentially retain the high image quality present in the image plane. As discussed below, the answer to this key question is not simple. We show in Section 7 of the present paper that, when the TIE-Hom retrieval is combined with an explicit deconvolution of the detector PSF, the image quality after the retrieval step becomes proportional to $\gamma^{1/2}$, being independent of the propagation distance and the detector PSF. The gain factor $\gamma^{1/2}$ is larger than $(\gamma / N_F)^{1/2}$ when $N_F > 1$, i.e. within the validity range of the TIE-Hom approximation. We then show in Section 8 that if CTF reconstruction is utilized, instead of the TIE-Hom reconstruction, in combination with the PSF deconvolution, the image quality gain can become proportional to $\gamma$, i.e. it can indeed reach the predicted ultimate upper limit of efficiency. This increase in the image quality gain, compared to the conventional TIE-Hom method, is subsequently confirmed in a numerical simulation example in Section 9. However, the CTF-Hom reconstruction needs to be regularized to mitigate the problem with zeros of the CTF at certain spatial frequencies which adversely affect the image reconstruction. The interplay between the maximization of the image quality gain in terms of the ratio of SNR to spatial resolution and the mitigation of the instability of the reconstruction near the CTF zeros is discussed both at the end of Section 8 and in the Conclusions.

## 2. Intensity distribution in in-line images

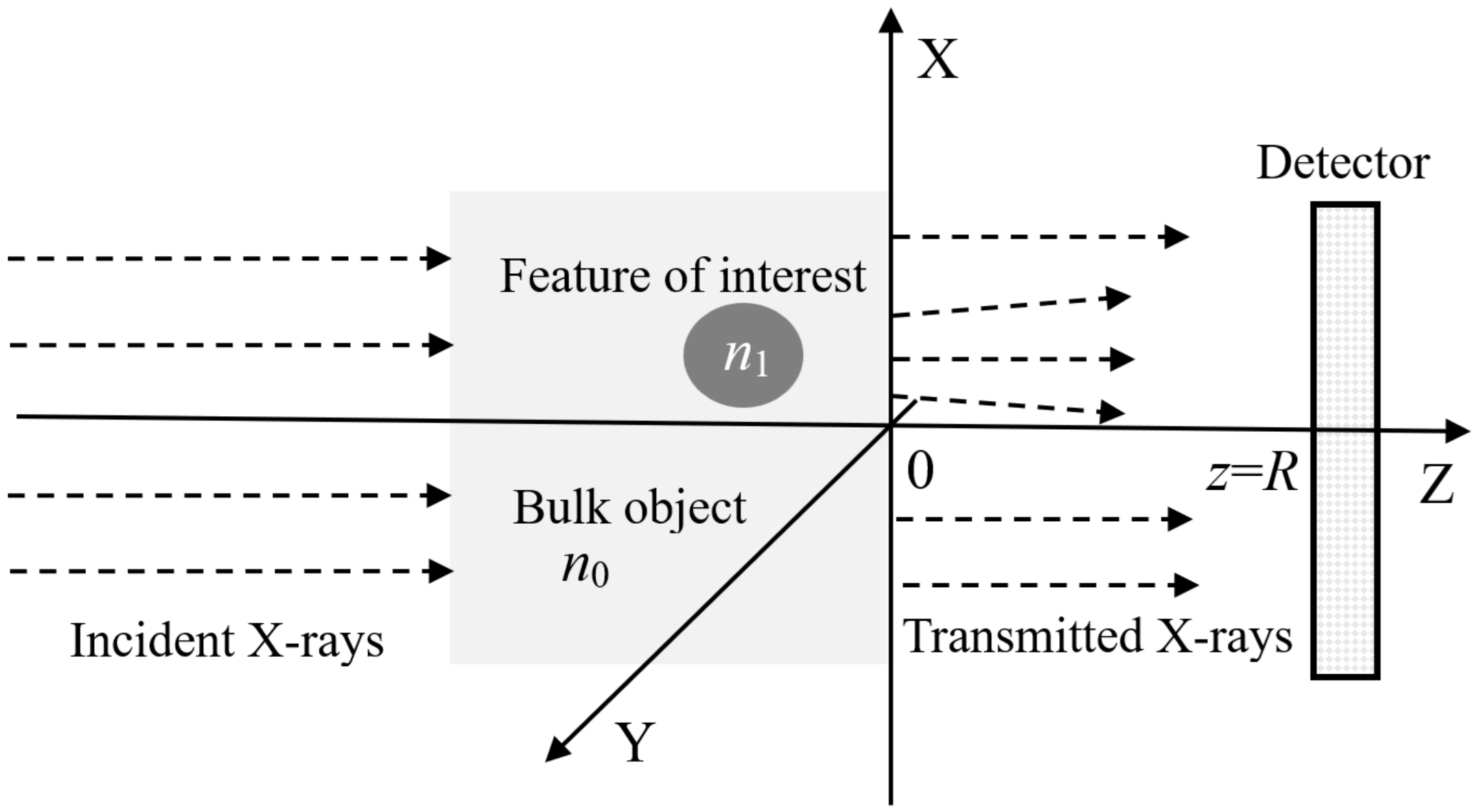

**Fig.1.** Imaging setup of propagation-based (in-line) imaging.

Let $\mathbf{r} = (x, y, z)$ and $\mathbf{r}_\perp = (x, y)$ be Cartesian coordinates in 3D and 2D space, respectively, with the vector length denoted $r = |\mathbf{r}|$ and $r_\perp = |\mathbf{r}_\perp|$, respectively. We assume that a small object feature with complex refractive index $n_1(\mathbf{r})$ is embedded in a large uniform "bulk" object with complex refractive index $n_0$, and that the distribution $\tilde{n}(\mathbf{r}) = n_1(\mathbf{r}) - n_0$ satisfies the homogeneity (monomorphicity) assumption (Paganin et al., 2002):

$$\tilde{n}(\mathbf{r}) = \beta(\mathbf{r})(i - \gamma), \quad (1)$$

where $\gamma$ is assumed to be constant inside the feature of interest[5]. The whole object is located near the optical axis $z$ immediately downstream of the "object plane" $z=0$ and is illuminated by a plane monochromatic X-ray wave with complex amplitude $U_{in}(\mathbf{r})=A\exp(ikz)$, where $A>0$ is the constant real amplitude, $k=2\pi/\lambda$ is the wavenumber and $\lambda$ is the X-ray wavelength (Fig.1). The intensity of the X-ray wave transmitted through the object is then equal to

$$|U|^2(\mathbf{r}_\perp,0)=A^2\exp[-B_0-B(\mathbf{r}_\perp)], \tag{2}$$

where $B_0=(4\pi/\lambda)\beta_0 T_0$, where $\beta_0=\operatorname{Im} n_0$, $T_0$ is the (uniform) thickness of the bulk object and

$$B(\mathbf{r}_\perp)=(4\pi/\lambda)\int\beta(\mathbf{r}_\perp,z)dz \tag{3}$$

is the projected absorption distribution of the feature of interest relative to the "background" absorption $B_0$. Note that eqs.(2) and (3) correspond to the Beer-Lambert absorption law.

Propagation of intensity of a transmitted wave, in the case of plane-wave illumination, can be described by the homogeneous Transport of Intensity equation (TIE-Hom) (Paganin *et al.*, 2002; Paganin, 2006):

$$|U|^2(\mathbf{r}_\perp,R)=(1-a^2\nabla_\perp^2)|U|^2(\mathbf{r}_\perp,0), \tag{4}$$

where $a^2=\gamma R\lambda/(4\pi)$. A sufficient condition for the validity of the TIE-Hom approximation eq.(4) is:

$$\Delta_{obj}^2>R\lambda, \tag{5}$$

[5] More generally, the sample's complex refractive index decrement $\tilde{n}(\mathbf{r})$ will have independent real and imaginary components. However, the "monomorphous object" assumption — which generalizes the concept of a "single-material object" that was mentioned earlier in the present paper — implies the real part of the decrement to be proportional to its imaginary part (because both such parts are each directly proportional to the local electron density in the sample, and thereby proportional to one another). This is the physical meaning of the "monomorphicity" assumption in eq.(1).

where $\Delta_{obj}$ is the size of the smallest detail in the distribution of $B(\mathbf{r}_\perp)$ (see Appendix for details). This validity condition effectively requires the spatial distribution of the complex refractive index in the imaged object to be sufficiently slowly varying (Teague, 1983)[6].

Let a position-sensitive X-ray detector have a quantum efficiency $\eta$ and a PSF $G(\mathbf{r}_\perp, \sigma_{det})$, where $G(\mathbf{r}_\perp, \sigma) = (2\pi\sigma^2)^{-1}\exp[-r_\perp^2/(2\sigma^2)]$ is a 2D Gaussian distribution. Then, according to eq.(2) and eq.(4), the mean of the detected photon fluence in the object and the detector planes will be, respectively,

$$\bar{I}_0(\mathbf{r}_\perp) = \bar{I}_{tr}\exp[-B(\mathbf{r}_\perp)] * G(\mathbf{r}_\perp, \sigma_{det}) \quad (6)$$

and

$$\bar{I}_R(\mathbf{r}_\perp) = \bar{I}_{tr}\exp[-B(\mathbf{r}_\perp)] * (1 - a^2\nabla_\perp^2)G(\mathbf{r}_\perp, \sigma_{det}), \quad (7)$$

where $\bar{I}_{tr} = \eta A^2 \exp(-B_0)$ is the mean of the detected "background" transmitted fluence in the object plane, $I_{tr}(\mathbf{r}_\perp)$, which corresponds to a detector with quantum efficiency $\eta$ and a Dirac delta-function PSF, $\delta_D(\mathbf{r}_\perp)$[7]. It is assumed that the spatial variation of the background transmitted fluence $I_{tr}(\mathbf{r}_\perp)$ is due only to the photon counting statistics, and the mean fluence transmitted through the bulk object, $\bar{I}_{tr}$, is the same at any point in the images. All the fluences are expressed in the units of inverse square length (corresponding to the number of photons per unit area).

For any detected fluence $I_z(\mathbf{r}_\perp)$, we will denote its power spectral density (Goodman, 2000) via $\tilde{I}_z(\boldsymbol{\rho}_\perp)$, where $\boldsymbol{\rho}_\perp = (u, v)$ are the 2D reciprocal space coordinates (spatial frequencies) dual to $\mathbf{r}_\perp = (x, y)$. We assume that the fluence $I_{tr}(\mathbf{r}_\perp)$ is Poisson-distributed and statistically independent at any two points $\mathbf{r}_\perp \neq \mathbf{r}'_\perp$. In this case, the noise distribution $\Delta I_{tr}(\mathbf{r}_\perp) = I_{tr}(\mathbf{r}_\perp) - \bar{I}_{tr}(\mathbf{r}_\perp)$ is white, i.e. the noise power spectral density (NPS) has the same value $\Delta\tilde{I}_\delta$ at any spatial frequency, and this constant value is equal to the mean fluence $\Delta\tilde{I}_{tr} = \bar{I}_{tr} = \eta A^2 \exp(-B_0)$ (Goodman, 2000). The variance of the noise distribution $\Delta I_{tr}(\mathbf{r}_\perp)$ is formally infinite, but it becomes finite after the convolution with the detector PSF (see below).

---

[6] As previously mentioned, the TIE (Teague, 1983) relates the longitudinal intensity rate of change of a paraxial monochromatic beam to both the intensity and phase in planes perpendicular to the optical axis. For uniform plane-wave illumination of a thin monomorphous object (e.g. a single-material object), eq.(1) allows phase to be formally eliminated from the TIE. Upon also making a first-order approximation to the longitudinal intensity derivative as a function of sample-to-detector distance, the TIE-Hom special case in eq.(4) is obtained without any further approximation.

[7] In the derivation of eq.(7), we have used the following simple identity: $\{(1 - a^2\nabla_\perp^2)\exp[-B(\mathbf{r}_\perp)]\} * G(\mathbf{r}_\perp, \sigma_{det}) = \exp[-B(\mathbf{r}_\perp)] * (1 - a^2\nabla_\perp^2)G(\mathbf{r}_\perp, \sigma_{det})$.

TIE-Hom retrieval acts on the detected fluence in the image plane $z = R$ and produces a "reconstructed" fluence, $I_{rec}(\mathbf{r}_\perp)$, in the object plane (Paganin et al., 2002). For the mean fluences, this can be written as

$$\bar{I}_{rec}(\mathbf{r}_\perp) = (1 - a^2\nabla_\perp^2)^{-1}\bar{I}_R(\mathbf{r}_\perp). \tag{8}$$

Comparing eq.(8) with eqs.(6-7), we see that the application of the TIE-Hom retrieval to the mean propagated image fluence transforms eq.(7) into eq.(6), i.e. it exactly reproduces the mean detected intensity in the object plane. However, in practice TIE-Hom retrieval is typically applied not to the mean image fluence, but to "raw" stochastic image fluences obtained in a single image exposure. Therefore, an important aspect of practical TIE-Hom retrieval is the assessment of the noise distribution, $\Delta I_{rec}(\mathbf{r}_\perp) = I_{rec}(\mathbf{r}_\perp) - \bar{I}_{rec}(\mathbf{r}_\perp)$, in the TIE-Hom retrieved images. This topic is considered in the next section.

## 3. Change of SNR in forward and inverse TIE-Hom imaging

We shall consider the noise distribution only in uniform "background" areas of the images, located away from the image of the feature of interest. Assuming that the detected fluences are spatially ergodic (Goodman, 2000) allows us to evaluate the image noise by averaging over the pixels in such uniform areas of the images. This would generally not be possible inside the image of the feature of interest, where the image contrast described by the term $\exp[-B(\mathbf{r}_\perp)] * G(\mathbf{r}_\perp, \sigma_{det})$ with $B(\mathbf{r}_\perp) > 0$ in eq.(6), may substantially contribute to the variance of image intensity, mixing up with the image noise. Noise propagation and SNR in 2D TIE-Hom imaging was studied in detail previously in (Nesterets et al., 2014; Gureyev *et* *al.*, 2025) and elsewhere. Here we only consider the facts that are relevant for the investigation of limits of PBI efficiency.

As the convolution with the detector PSF and the forward TIE-Hom operator in the parallel-beam case both conserve the mean photon fluence, the mean fluence in the background parts of the images, $\bar{I}_{z,bg}(\mathbf{r}_\perp)$, is the same at any point $\mathbf{r}_\perp$ in the object and detector planes in the plane incident wave case: $\bar{I}_{0,bg}(\mathbf{r}_\perp) = \bar{I}_{R,bg}(\mathbf{r}_\perp) = \bar{I}_{tr} * G(\mathbf{r}_\perp, \sigma_{det}) = \bar{I}_{tr}$. The corresponding noise distribution is equal to $\Delta I_{0,bg}(\mathbf{r}_\perp) = \Delta I_{R,bg}(\mathbf{r}_\perp) = \Delta I_{tr}(\mathbf{r}_\perp) * G(\mathbf{r}_\perp, \sigma_{det})$. It is known (Goodman, 2000) that the power spectral density of a linearly-filtered stochastic distribution is equal to the power spectral density of the original distribution multiplied by the square modulus of the Fourier transform of the filter function (Goodman, 2000). Therefore, $\Delta\tilde{I}_{0,bg}(\boldsymbol{\rho}_\perp) = \Delta\tilde{I}_{R,bg}(\boldsymbol{\rho}_\perp) = \Delta\tilde{I}_{tr} \,|\, \hat{G}(\boldsymbol{\rho}_\perp, \sigma_{det}) \,|^2 = \bar{I}_{tr} \,|\, \hat{G}(\boldsymbol{\rho}_\perp, \sigma_{det}) \,|^2$, where the overhead hat symbol denotes the 2D Fourier transform $\hat{f}(\boldsymbol{\rho}_\perp) \equiv \iint \exp(-i2\pi\boldsymbol{\rho}_\perp \bullet \mathbf{r}_\perp) f(\mathbf{r}_\perp) d\mathbf{r}_\perp$. The variance of a stochastic spatial distribution with zero mean is equal to the integral of its power spectral density (Goodman, 2000). Using Parseval's theorem and the fact that $\iint |\, G(\mathbf{r}_\perp, \sigma) \,|^2 \, d\mathbf{r}_\perp = 1/(4\pi\sigma^2) = 1/\Delta^2$, where $\Delta \equiv 2\sqrt{\pi}\sigma$, we can find the variance of noise in the object and detector planes as integrals of the corresponding power spectra:

$$\mathrm{Var}[I_0] = \mathrm{Var}[I_R] = \tilde{I}_{tr} \iint |\, \hat{G}(\boldsymbol{\rho}_\perp, \sigma_{det}) \,|^2 \, d\boldsymbol{\rho}_\perp = \bar{I}_{tr} / \Delta_{det}^2 . \tag{9}$$

This result has a clear physical implication: filtering the photon fluence with a wide detector PSF decreases the noise variance in the detected fluence, at the expense of the spatial resolution. This trade-off constitutes the essence of the noise-resolution uncertainty (NRU) principle (Gureyev et al., 2014, 2016).

The propagation of noise in TIE-Hom retrieval is determined by the action of the TIE-Hom retrieval operator on the noise in the detected image fluence at $z = R$:

$$\Delta I_{rec,bg}(\mathbf{r}_\perp) = (1 - a^2 \nabla_\perp^2)^{-1} \Delta I_{R,bg}(\mathbf{r}_\perp) . \tag{10}$$

Taking into account that $\mathrm{Var}[I_{rec}] = \iint \Delta\tilde{I}_{rec,bg}(\boldsymbol{\rho}_\perp) d\boldsymbol{\rho}_\perp$ and proceeding as above, we obtain:

$$\mathrm{Var}[I_{rec}] = \Delta\tilde{I}_{tr} \iint |\hat{G}(\boldsymbol{\rho}_\perp, \sigma_{det})|^2 |\hat{T}_{inv}(\boldsymbol{\rho}_\perp, a)|^2 \, d\boldsymbol{\rho}_\perp \,, \tag{11}$$

where $\hat{T}_{inv}(\boldsymbol{\rho}_\perp, a) = (1 + 4\pi^2 a^2 \rho_\perp^2)^{-1}$ is the Fourier transform of the filter function $T_{inv}(\mathbf{r}_\perp, a) = (1 - a^2\nabla_\perp^2)^{-1}\delta_D(\mathbf{r}_\perp)$ (Nesterets et al., 2014). From eq.(11), we obtain $\mathrm{Var}[I_{rec}] = \bar{I}_{tr} \iint \exp(-4\pi^2\sigma_{det}^2\rho^2)(1+4\pi^2a^2\rho^2)^{-2} d\boldsymbol{\rho}_\perp$. Introducing new variables $\boldsymbol{\rho}'_\perp = 2\pi\sigma_{det}\boldsymbol{\rho}_\perp$, switching to polar coordinates $(\rho', \theta)$, integrating over $\theta$ and finally replacing $\rho'^2 = t$, we obtain

$$\begin{aligned} \mathrm{Var}[I_{rec}] &= \frac{\bar{I}_{tr}}{4\pi\sigma_{det}^2} \int_0^\infty \frac{\exp(-t)}{(1 + t\gamma / N_F)^2} dt = \\ & (\bar{I}_{tr} / \Delta_{det}^2)(N_F / \gamma)[1 - (N_F / \gamma)\exp(N_F / \gamma) E_1(N_F / \gamma)] = \\ & (\bar{I}_{tr} / \Delta_{det}^2)\{(N_F / \gamma) + (N_F / \gamma)^2[\ln(N_F / \gamma) + \gamma_{EM}] + O((N_F / \gamma)^3 \ln(N_F / \gamma))\}, \end{aligned} \tag{12}$$

where $N_F = \Delta_{det}^2 / (\lambda R)$ is the Fresnel number associated with the detector resolution, $E_1(w) = \int_w^\infty \exp(-t)t^{-1}dt$ is a standard special function ("exponential integral") (Abramowitz & Stegun, 1964), $\gamma_{EM} \cong 0.57721$ is the Euler-Mascheroni constant, and the last approximation in eq.(12) is based on an asymptotic expansion at $\gamma / N_F \to \infty$.

The mean fluence $\bar{I}_{tr}$ in background areas of images in the object and detector planes is equal to the mean number of photons detected in a pixel, divided by the pixel area. This mean fluence is unaffected by the PSF of the detector (the quantum efficiency of the detector is taken into account separately). The forward and inverse TIE-Hom operators also do not change the mean fluence. Defining the corresponding SNR in the usual manner as the ratio of the mean value of image fluence to its standard deviation, $\mathrm{SNR}[I] = \bar{I} / \mathrm{Var}^{1/2}[I]$, we obtain from eqs.(9,12):

$$\mathrm{SNR}^2[I_R] = \mathrm{SNR}^2[I_0] = \bar{I}_{tr}\Delta_{det}^2 \,, \tag{13a}$$

$$\mathrm{SNR}^2[I_{rec}] \cong \mathrm{SNR}^2[I_0]\,[(\gamma / N_F) + \ln(\gamma / N_F) - \gamma_{EM}] \,, \tag{13b}$$

where eq.(13b) was obtained with the use of the approximation $(\gamma / N_F) / \{1+(N_F / \gamma)[\ln(N_F / \gamma)+\gamma_{EM}]\} \cong (\gamma / N_F)\{1-(N_F / \gamma)[\ln(N_F / \gamma)+\gamma_{EM}]\}$ valid in the case of $\gamma / N_{F,det} >> 1$. In particular, in the case of $\gamma / N_F >> 1$, we obtain:

$$G_{2D} \equiv \mathrm{SNR}[I_{rec}] / \mathrm{SNR}[I_0] \cong (\gamma / N_F)^{1/2}, \tag{14}$$

which agrees with the previously reported results (Nesterets et al., 2014; Gureyev *et al.*, 2025). The "gain factor" $G_{2D}$ quantifies the gain in the image SNR after coherent free-space propagation and TIE-Hom retrieval, in comparison with the "contact" (ABI) image in the plane $z = 0$. This gain, associated with the optical physics of image-formation "encoding" in X-ray PBI for homogeneous samples, constitutes an example of the much more general "noisy-channel coding theorem" mentioned earlier in our paper.

It is important to note that while the factor $(\gamma / N_F)^{1/2}$ can in principle become infinitely large when $N_F \to 0$ (e.g. when propagation distance $R$ becomes very large), this gain is limited by the range of validity of the TIE-Hom approximation. The TIE-Hom retrieval correctly reconstructs the intensity in the object plane only when eq.(5) holds. Therefore, the upper limit of the gain in eq.(14) satisfies the inequality $(\gamma / N_F)^{1/2} \leq (\gamma R \lambda / \Delta_{obj}^2)^{1/2} < \gamma^{1/2}$. This outcome is in agreement with the general idea discussed in the introduction that the gain in phase-contrast imaging quality, relative to the quality of the corresponding absorption imaging, is determined by the parameter $\gamma$.

In the next section we shall investigate how the changes in the SNR in forward and inverse TIE-Hom imaging relate to changes in the spatial resolution. This topic is closely linked with the previously mentioned NRU (Gureyev et al., 2014, 2016).

## 4. Spatial resolution and NRU in forward and inverse TIE-Hom imaging

We define the spatial resolution of an imaging system as the width of its PSF, i.e. the width of the response of the system to a delta-function-like input. Regarding the ways to define the width of a 2D function, we will use the following definition:

$$\Delta[F] = |\iint F(\mathbf{r}_\perp) d\mathbf{r}_\perp | / (\iint | F(\mathbf{r}_\perp)|^2 \, d\mathbf{r}_\perp)^{1/2} . \qquad (15)$$

It can be easily verified that for a function $F_{rect}(\mathbf{r}_\perp, h)$, that is equal to a constant value $A \neq 0$ within the square region $\{-h/2 \leq x \leq h/2, -h/2 \leq y \leq h/2\}$ and is equal to zero outside this region, we obtain $\Delta[F_{rect}(\mathbf{r}_\perp, h)] = Ah^2 / (A^2 h^2)^{1/2} = h$, which obviously coincides with the conventional notion of the width of a square. In the case of a Gaussian function, we obtain $\Delta[G(\mathbf{r}_\perp, \sigma)] = 2\sqrt{\pi}\sigma \cong 3.545\sigma$. This coincides with the notation that was used above for the Gaussian detector PSF, $\Delta_{det} = 2\sqrt{\pi}\sigma_{det}$. This definition also agrees in principle with the usual understanding of the width of a Gaussian as a multiple of its standard deviation, as e.g. in the usual definition of the full width at half-maximum: $\text{FWHM} = 2\sqrt{2\ln 2}\,\sigma \cong 2.355\sigma$. Our choice of the factor $2\sqrt{\pi}$, rather than $2\sqrt{2\ln 2}$, is determined by notational convenience: it simplifies the form of many relevant mathematical expressions below.

Definitions of function width similar to eq.(15) were previously used e.g. in (Mandel, 1959; Mandel & Wolf, 1962; Gureyev et al., 2016). However, in these earlier studies the numerator in eq.(15) was either implicitly assumed to be equal to unity or had the form $\iint | F(\mathbf{r}_\perp) | d\mathbf{r}_\perp$, rather than $| \iint F(\mathbf{r}_\perp) d\mathbf{r}_\perp |$. As such definitions were typically used in conjunction with non-negative functions $F(\mathbf{r}_\perp)$ (which in most cases also had the integral value of one), they were equivalent to eq.(15). However, in the context of free-space propagation, the PSF of the imaging system can develop negative lobes, corresponding to Fresnel fringes. Physically, this PSF describes the process of re-distribution of photons from one area of the image to an adjacent one. In such cases, as shown below, it is important to use the definition of the width in the form of eq.(15), as the photon conservation law implies that the integral of the PSF remains the same at any propagation distance, while the integral of the absolute value of the PSF is not conserved.

Consider the case of a spatially uniform and uncorrelated photon flux (Goodman, 2000) incident on an ideal photon-counting detector with a single-pixel PSF and square pixels with side length $\Delta_{pix}$. Let $\bar{n}_{pix}$ be the mean number of registered photons in a pixel. If we bin every four adjacent pixels (in a two by two configuration, i.e. taking two adjacent pixels in a row and two pixels in a column), thus creating bigger effective pixels with the side length $2\Delta_{pix}$, then each one of the new pixels will have $4\bar{n}_{pix}$ detected photons on average. This process transforms the initial image with the spatial resolution $\Delta_{pix}$ and SNR equal to $\bar{n}_{pix}^{1/2}$ (assuming Poisson statistics) into a new image with a spatial resolution of $2\Delta_{pix}$ and an SNR equal to $(4\bar{n}_{pix})^{1/2} = 2\bar{n}_{pix}^{1/2}$. Note that the ratio of the SNR to spatial resolution remains unchanged in this transformation. The NRU is a generalization of this simple phenomenon, which states that the ratio of the SNR to the spatial resolution remains constant in a broad class of intensity-linear transformations of images (Gureyev et al., 2014, 2016). In the present setting, this result can be expressed as

$$\frac{\mathrm{SNR}[I_0 * F]}{\Delta[P * F]} = \frac{\mathrm{SNR}[I_0]}{\Delta[P]}, \tag{16}$$

where $I_0(\mathbf{r}_\perp)$ is an original image, $F(\mathbf{r}_\perp)$ is the filter function and $P(\mathbf{r}_\perp)$ is the PSF corresponding to the original image. Here we are concerned with special cases of the general NRU, eq.(16), involving Gaussian detector PSFs $P(\mathbf{r}_\perp) = G(\mathbf{r}_\perp, \sigma_{det})$ and filter functions corresponding to forward and inverse TIE-Hom, $F(\mathbf{r}_\perp) = T_{frw}(\mathbf{r}_\perp, a)$ and $F(\mathbf{r}_\perp) = T_{inv}(\mathbf{r}_\perp, a)$. We have already analysed the behaviour of the SNR after the filtering operation corresponding to the forward and inverse TIE-Hom retrieval, with the results given by eqs.(13). Now we shall consider the behaviour of the spatial resolution in the forward and inverse stages of TIE-Hom imaging. It is important to keep in mind that the NRU in the form of eq.(16) was proved only for non-negative PSFs and filter functions (or, equivalently, for absolute values of functions) (de Hoog et al., 2014; Gureyev et al., 2014, 2016).

Let $P_0(\mathbf{r}_\perp)$ and $P_R(\mathbf{r}_\perp)$ be the PSFs in the object and the detector planes, respectively. Firstly, it is easy to see that $P_0(\mathbf{r}_\perp) = G(\mathbf{r}_\perp, \sigma_{det})$ and hence $\Delta[P_0] = \Delta[G(\mathbf{r}_\perp, \sigma_{det})] = \Delta_{det}$. Next, according to eqs.(6) and (7), the PSF changes from $P_0(x, y)$ to $P_R(\mathbf{r}_\perp) = (1 - a^2\nabla_\perp^2)G(\mathbf{r}_\perp, \sigma_{det}) = (T_{frw} * G_{det})(\mathbf{r}_\perp)$ upon free-space propagation over distance $R$. We can calculate the change of the width of the PSF in that transformation. Using integration by parts, it is straightforward to verify that the forward TIE-Hom operator leaves the integral of the PSF unchanged: $\iint (1 - a^2\nabla_\perp^2)P(\mathbf{r}_\perp)d\mathbf{r}_\perp = \iint P(\mathbf{r}_\perp)d\mathbf{r}_\perp = 1$. We can use Parseval's theorem to calculate the second integral norm: $\|(1-a^2\nabla_\perp^2)P\|_2 = \|(1+4\pi^2a^2\rho^2)\hat{P}(\boldsymbol{\rho}_\perp)\|_2$, where $\|F\|_2 \equiv (\iint |F(\mathbf{r}_\perp)|^2\, d\mathbf{r}_\perp)^{1/2}$. Next, we proceed as in the derivation of eq.(12) above to calculate the following integral: $\|(1-a^2\nabla_\perp^2)G(\mathbf{r}_\perp,\sigma_{det})\|_2^2 =$
$\iint (1+4\pi^2a^2\rho^2)^2 \exp(-4\pi^2\sigma_{det}^2\rho^2)d\boldsymbol{\rho}_\perp = \Delta_{det}^{-2}[1+2(\gamma/N_F)+2(\gamma/N_F)^2]$. Therefore,

$$\Delta^2[P_R] = \frac{1}{\|(1-a^2\nabla_\perp^2)G(\mathbf{r}_\perp,\sigma_{det})\|_2^2} = \frac{\Delta^2[P_0]}{1+2(\gamma/N_F)+2(\gamma/N_F)^2}. \tag{17a}$$

The effect of the TIE-Hom retrieval on the spatial resolution can be evaluated by taking into account that $P_{rec} = (1-a^2\nabla_\perp^2)^{-1}P_R = (1-a^2\nabla_\perp^2)^{-1}(1-a^2\nabla_\perp^2)P_0 = P_0$. Therefore,

$$\Delta^2[P_{rec}] = \Delta^2[P_0] = \Delta_{det}^2. \tag{17b}$$

Now consider the squared ratios of the SNR to spatial resolution in the forward and inverse TIE-Hom imaging, which can be evaluated using eqs.(13a-b) and (17a-b). First consider the case of forward TIE-Hom imaging. As $\mathrm{SNR}^2[I_0]/\Delta^2[P_0] = \bar{I}_{tr}$ and
$\mathrm{SNR}^2[I_R]/\Delta^2[P_R] = \bar{I}_{tr}[1+2(\gamma/N_F)+2(\gamma/N_F)^2]$, we find that

$$\frac{\mathrm{SNR}^2[I_R]}{\Delta^2[P_R]} = \frac{\mathrm{SNR}^2[I_0]}{\Delta^2[P_0]}[1+2(\gamma/N_F)+2(\gamma/N_F)^2] \cong \frac{\mathrm{SNR}^2[I_0]}{\Delta^2[P_0]}2\left(\frac{\gamma}{N_F}\right)^2, \tag{18a}$$

where the last approximation was obtained under the condition $\gamma/N_F >> 1$.

Similarly, considering the TIE-Hom retrieval step, we can also obtain the following relationships from eqs.(13a-b) and (17a-b):
$\mathrm{SNR}^2[I_{rec}]/\Delta^2[P_{rec}] \cong \bar{I}_{tr}[(\gamma/N_F)+\ln(\gamma/N_F)-\gamma_{EM}]$. Therefore,

$$\frac{\mathrm{SNR}^2[I_{rec}]}{\Delta^2[P_{rec}]} \cong \frac{\mathrm{SNR}^2[I_R]}{\Delta^2[P_R]}\frac{(\gamma/N_F)+\ln(\gamma/N_F)-\gamma_{EM}}{1+2(\gamma/N_F)+2(\gamma/N_F)^2} \cong \frac{\mathrm{SNR}^2[I_R]}{\Delta^2[P_R]}\frac{N_F}{2\gamma}, \tag{18b}$$

with the last approximation obtained under the condition $\gamma/N_F >> 1$.

Finally, we can obtain the relationship between the ratios of SNR to spatial resolution in the reconstructed images, in comparison with similar ratios in the contact images collected in the plane $z = 0$:

$$\frac{\mathrm{SNR}^2[I_{rec}]}{\Delta^2[P_{rec}]} \cong \frac{\mathrm{SNR}^2[I_0]}{\Delta^2[P_0]}[(\gamma / N_F) + \ln(\gamma / N_F) - \gamma_{EM}] \cong \frac{\mathrm{SNR}^2[I_0]}{\Delta^2[P_0]} \frac{\gamma}{N_F} \ . \tag{19}$$

Equation (19) corresponds to a typical scenario in experimental TIE-Hom imaging. The right-hand side of eq.(19) is proportional to the square of the 2D TIE-Hom gain factor $G_{2D}$ from eq.(14), which agrees with previously published results (Gureyev *et al.*, 2025).

## 5. Violations of NRU in forward and inverse TIE-Hom imaging

Equations (18a) and (18b) appear to violate the NRU, namely eq.(16). In particular, equation (18a) shows that the NRU does not hold in the forward TIE-Hom imaging. This violation is a "beneficial" one, in the sense that the ratio of SNR to spatial resolution becomes larger after the coherent free-space propagation. The violation of the NRU is caused here by the fact that the spatial resolution is improved in the forward propagation before the image is registered (essentially, at the level of complex amplitudes), while the SNR in the registered image remains the same as in the object plane because: (i) the mean number of photons is conserved in free-space imaging, and (ii) the noise is filtered by the "post-registration" PSF, which is the same detector PSF as in the object plane, and is not affected by the forward propagation (Gureyev *et al.*, 2017). In other words, the spatial resolution and the noise in the propagated image are affected by different filter functions.

Equation (18b) indicates that the NRU is violated in the TIE-Hom retrieval as well. It appears that this result has not been previously described in the published literature. Unlike the case with eq.(18a), this violation of the NRU is "detrimental", because the ratio of the SNR to spatial resolution decreases after the TIE-Hom retrieval. More specifically, while both the SNR and the width of the PSF become larger, the deterioration in the spatial resolution is stronger than the gain in the SNR. It has been argued previously (Gureyev *et al.*, 2017, 2025) that the ratio of SNR to spatial resolution remains unchanged in the TIE-Hom retrieval.

Equation (18b) demonstrates that this is not the case. The reasons for NRU violation in eqs.(18a) and (18b) are different. In the case of eqs.(18a), the violation of the NRU is caused by the fact that the noise is filtered by a function (the detector PSF $G_{det}(\mathbf{r}_\perp)$ ) that is different from the one that determines the spatial resolution in the imaged object (the convolution of the detector PSF with the forward TIE-Hom propagation, $(T_{frw} * G_{det})(\mathbf{r}_\perp)$ ). In the case of eq.(18b), both the noise and the spatial resolution are affected by the same filter function $T_{inv}(\mathbf{r}_\perp, a) = (1 - a^2\nabla_\perp^2)^{-1}\delta_D(\mathbf{r}_\perp)$ in the course of TIE-Hom retrieval. However, the presence of negative values of the PSF $P_R(\mathbf{r}_\perp) = (1 - a^2\nabla_\perp^2)G(\mathbf{r}_\perp, \sigma_{det})$ violates the mathematical conditions required for the NRU to hold (de Hoog et al., 2014; Gureyev et al., 2016). As a result, the ratio of SNR to spatial resolution does not remain constant, but deteriorates approximately in proportion to $(2\gamma / N_{F,det})^{-1/2} < 1$ after the TIE-Hom retrieval (see eq.(18b)). However, since the ratio of SNR to spatial resolution was larger by the factor $2^{1/2}\gamma / N_{F,det}$ in the image plane compared to the object plane (eq.(18a)), the combined effect of the forward propagation and retrieval steps in the TIE-Hom imaging is an increase of the SNR to resolution ratio by the factor $(\gamma / N_{F,det})^{1/2} >> 1$, compared to the same ratio in contact imaging at $z = 0$ (see eq.(19)).

Table 1 contains a summary of the above results, including a restatement in terms of the two-dimensional "intrinsic imaging quality characteristic" (Gureyev et al., 2014, 2016), $\mathrm{Q}_{S,2D}^2 = \mathrm{SNR}^2 / (\bar{I}_{in}\,\Delta^2)$, where $\bar{I}_{in} = A^2$ is the mean incident photon fluence. The intrinsic imaging quality characteristic is a measure of Shannon information that the imaging system is capable of extracting on average from a single incident photon. As can be seen from the last row of Table 1, this imaging characteristic is always proportional to the quantum efficiency of the detector and inversely proportional to the X-ray absorption in the bulk of the sample. These two properties are quite straightforward and easy to appreciate. On the other hand, the increase of the intrinsic imaging quality in proportion to the gain factor $G_{2D}$ that can be observed in the two last cells in the bottom row of Table 1, is non-trivial and represents the key beneficial property of PBI (Gureyev et al., 2017, 2025). As noted above, the image quality gain is larger in the image plane before the TIE-Hom retrieval, than in the object plane after the retrieval. It was also noted earlier that the image quality gain is limited by the range of validity of the TIE-Hom approximation, eq.(5). This implies that the overall gain in the intrinsic imaging quality characteristic in TIE-Hom imaging is limited from above as $Q^2_{S,2D} = \eta e^{-B_0}(\gamma / N_F) < \eta e^{-B_0}\gamma$ .

**Table 1.** SNR and spatial resolution in 2D distributions of photon fluence in flat areas of images in PBI.

| | Object plane (original) | Image plane | Object plane (retrieved) |
|---|---|---|---|
| SNR$^2[I]$ in flat fields | $\bar{I}_{tr}\Delta^2_{det}$ | $\bar{I}_{tr}\Delta^2_{det}$ | $\bar{I}_{tr}\Delta^2_{det}(\gamma / N_F)$ |
| $\Delta^2$ | $\Delta^2_{det}$ | $\dfrac{\Delta^2_{det}}{2(\gamma / N_F)^2}$ | $\Delta^2_{det}$ |
| $Q^2_{S,2D}$ | $\eta e^{-B_0}$ | $2\eta e^{-B_0}(\gamma / N_F)^2$ | $\eta e^{-B_0}(\gamma / N_F)$ |

## 6. Propagation-based CT with TIE-Hom retrieval

The NPS of a reconstructed 3D distribution of the imaginary part, $\beta$, of the refractive index of the sample, obtained using PB-CT with TIE-Hom retrieval, is equal to

$W_{R,\beta}(u,v,\theta) = \dfrac{\pi u |G(u,v,\sigma_{det})|^2}{(2k)^2 \bar{I}_{tr,CT}[1+4\pi^2 a^2(u^2+v^2)]^2}$ (Nesterets et al., 2014; Gureyev *et al.*, 2025),

where $\theta$ is the CT view angle, $\bar{I}_{tr,CT} \equiv M_a \bar{I}_{tr}$ is the mean transmitted photon fluence detected in a unit area during the CT scan, $M_a$ is the number of projections (view angles) in the CT scan, $u$ is the radial coordinate in planes orthogonal to the CT rotation axis and $v$ is the coordinate along the rotation axis. Correspondingly, in the case of a detector with a Gaussian PSF, the noise variance in the reconstructed distribution of $\beta$ is equal to the integral of $W_{R,\beta}(u,v,\theta)$, i.e. $\mathrm{Var}[\beta_{rec}] = \dfrac{2\pi^2}{(2k)^2\bar{I}_{tr,CT}} \int_{-\infty}^{\infty}\int_{0}^{\infty} \dfrac{u^2 \exp[-4\pi^2\sigma^2_{det}(u^2+v^2)]dudv}{[1+4\pi^2a^2(u^2+v^2)]^2}$. This integral can be evaluated in a similar manner to the derivation of eq.(12) above, with the result

$\mathrm{Var}[\beta_{rec}] = \dfrac{\pi[(b^3+b^2)\exp(b)E_1(b) - b^2]}{2(2k)^2\bar{I}_{tr,CT}\Delta^4_{det}} \cong \dfrac{\pi[b^2(\ln b^{-1} - \gamma_{EM} - 1)]}{2(2k)^2\bar{I}_{tr,CT}\Delta^4_{det}}$, where $b \equiv N_F / \gamma$ and the

last approximation was made for the case $b \ll 1$. As the reconstructed mean "signal" in PB-CT in this case is equal to $\bar{\beta}_{rec}$, we obtain

$$\mathrm{SNR}^2[\beta_{rec}] \cong (2/\pi)\bar{\mu}^2 \bar{I}_{tr,CT}\, \Delta_{det}^4 (\gamma / N_F)^2 / [\ln(\gamma / N_F) - \gamma_{EM} - 1], \tag{20}$$

where $\bar{\mu} = 2k\bar{\beta}_{rec}$ and we ignored terms of the order $b^3 \ln(1/b)$ and smaller. In the case of contact imaging ($a$ = 0), we can similarly derive that $\mathrm{Var}[\beta_0] = \frac{2\pi^2}{(2k)^2 \bar{I}_{tr,CT}} \int_{-\infty}^{\infty}\int_{0}^{\infty} u^2 \exp[-4\pi^2\sigma_{det}^2(u^2+v^2)]dudv = \frac{\pi}{2(2k)^2 \bar{I}_{tr,CT}\Delta_{det}^4}$, which implies that

$$\mathrm{SNR}^2[\beta_0] = (2/\pi)\bar{\mu}^2 \bar{I}_{tr,CT}\Delta_{det}^4. \tag{21}$$

From eqs.(20)-(21) we find that the 3D gain factor in PB-CT is equal to

$$G_{3D} \equiv \frac{\mathrm{SNR}[\beta_{rec}]}{\mathrm{SNR}[\beta_0]} \cong \frac{\gamma / N_F}{[\ln(\gamma / N_F) - \gamma_{EM} - 1]^{1/2}}, \tag{22}$$

which agrees with the result reported earlier in (Nesterets et al., 2014).

Provided that the Nyquist sampling conditions are satisfied, i.e. a sufficient number of projections has been collected at equispaced view angles over 180 degrees of rotation of the sample, the spatial resolution after the PB-CT reconstruction is approximately equal to $\Delta[P_{rec}] = \Delta_{det}$ (Natterer, 2001). If we also normalize the appropriate ratio of the SNR to spatial resolution by the 3D incident photon density (Gureyev *et al.*, 2025), we obtain the following expression for the square of the 3D intrinsic imaging quality characteristic:

$$\mathrm{Q}_{S,3D}^2[\beta_{rec}] \equiv \frac{\mathrm{SNR}^2[\beta_{rec}]}{\bar{I}_{in,3D}\,\Delta^3[P_{rec}]} \cong \frac{\eta \exp(-B_0)(\bar{\mu}L)^2}{M_R} \frac{(\gamma / N_F)^2}{\ln(\gamma / N_F) - \gamma_{EM} - 1}. \tag{23a}$$

Above we have taken into account the fact that $\bar{I}_{in,3D} = N_{3D} / V = N_{2D} M_a / (\Omega L) = \bar{I}_{in,2D} M_a / L = \bar{I}_{tr,CT} / [L\eta \exp(-B_0)]$. Moreover, (i) $N_{3D}$ and $N_{2D}$ are the total numbers of incident photons in the CT scans and in one projection, respectively, (ii) $\bar{I}_{in,2D} = \bar{I}_{tr} / [\eta \exp(-B_0)]$, and (iii) $L = V / \Omega$ is the effective depth of the sample defined as the ratio of the sample volume, $V$, and the area of the irradiated front surface, $\Omega$. We have also introduced the notation $M_R = (\pi/2)L/\Delta_{det} = (\pi/2)^2 R_C / \Delta_{det}$ for the quantity equal to $(\pi/2)^2$ times the number of spatial resolution units in the radius $R_C$ of the CT reconstruction volume (which has a cylindrical shape). The parameter $M_R$ is an indicator of the degree of mathematical ill-posedness of the CT reconstruction (Natterer, 2001; Gureyev *et al.*, 2025).

Similarly,

$$Q^2_{S,3D}[\beta_0] \equiv \frac{\text{SNR}^2[\beta_0]}{\bar{I}_{in,3D}\,\Delta^3[P_0]} = \frac{\eta \exp(-B_0)(\bar{\mu}L)^2}{M_R}. \tag{23b}$$

In the same way as done in the 2D TIE-Hom imaging case above, we shall also compare the 3D imaging quality gain obtained after the free-space propagation followed by the TIE-Hom retrieval with a similar gain obtained in the image plane $z = R$ without the retrieval step. If the CT reconstruction is performed using the projections collected in the plane $z = R$, without the TIE-Hom retrieval, the SNR will be the same as in eq.(21), because the mean number of photons and the NPS will be the same as in the contact CT case. As for the spatial resolution in the CT reconstruction in the image plane, it is commensurate with the spatial resolution in the projections. According to eq.(17a), that spatial resolution is equal to $\Delta[P_R] = \Delta_{det}/[1+2(\gamma/N_F)+2(\gamma/N_F)^2]^{1/2}$. Note that due to this finer spatial resolution in PB-CT projections, it may be necessary to increase the number of projection angles in order to satisfy the Nyquist sampling conditions and achieve the same spatial resolution in the CT-reconstructed images as in the individual projections. However, this requirement does not affect the value of the 3D intrinsic imaging quality, as the number of projections is the same in its numerator and denominator and, hence, is cancelled out from the following expression:

$$Q^2_{S,3D}[\beta_R] \equiv \frac{\text{SNR}^2[\beta_R]}{\bar{I}_{in,3D}\,\Delta^3[P_R]} \cong \frac{\eta \exp(-B_0)(\bar{\mu}L)^2}{M_R} 2^{3/2}(\gamma/N_F)^3. \tag{23c}$$

It follows from eqs.(23a-b), that the gain in intrinsic imaging quality of PB-CT in comparison with absorption ("contact") CT at the same incident fluence and spatial resolution is equal to

$$Q_{S,3D}[\beta_{rec}]/Q_{S,3D}[\beta_0] = G_{3D} \cong (\gamma/N_F)/[\ln(\gamma/N_F) - \gamma_{EM} - 1]^{1/2}. \tag{24}$$

This gain factor in 3D case is higher than the one in 2D planar imaging given by eq.(14). Similar results were reported previously in (Nesterets et al., 2014; Gureyev *et al.*, 2025). The ultimate upper limit of the gain factor $G_{3D}$ is determined by eq.(5) and is asymptotically equal to $\gamma \ln^{-1/2}(\gamma)$ when $\ln\gamma >> 1$. On the other hand, comparing eqs.(23a) and (23c), we see that, similarly to the 2D case, the intrinsic imaging quality in PB-CT is actually higher in the image plane before the TIE-Hom retrieval, than it is in the object plane after the TIE-Hom retrieval: $\mathrm{Q}_{S,3D}[\beta_R] / \mathrm{Q}_{S,3D}[\beta_{rec}] \cong 2^{3/4}\{(\gamma / N_F)[\ln(\gamma / N_F) - \gamma_{EM} - 1]\}^{1/2}$.

**Table 2.** SNR and spatial resolution in the reconstructed 3D distributions of $\beta(\mathbf{r})$ in PB-CT imaging in the case $\ln(\gamma / N_F) >> 1$.

| | Object plane (absorption CT) | Image plane | Object plane (PB-CT) |
|---|---|---|---|
| $\mathrm{SNR}^2[\beta]$ in flat areas | $\frac{2}{\pi}\bar{I}_{tr,CT}\bar{\mu}^2\Delta_{det}^4$ | $\frac{2}{\pi}\bar{I}_{tr,CT}\bar{\mu}^2\Delta_{det}^4$ | $\frac{2}{\pi}\bar{I}_{tr,CT}\bar{\mu}^2\Delta_{det}^4 \frac{(\gamma / N_F)^2}{\ln(\gamma / N_F)}$ |
| $\Delta^3$ | $\Delta_{det}^3$ | $\frac{\Delta_{det}^3}{2^{3/2}(\gamma / N_F)^3}$ | $\Delta_{det}^3$ |
| $\mathrm{Q}_{S,3D}^2[\beta]$ | $\frac{\eta e^{-B_0}(\bar{\mu}L)^2}{M_R}$ | $\frac{2^{3/2}\eta e^{-B_0}(\bar{\mu}L)^2}{M_R}\left(\frac{\gamma}{N_F}\right)^3$ | $\frac{\eta e^{-B_0}(\bar{\mu}L)^2}{M_R}\frac{(\gamma / N_F)^2}{\ln(\gamma / N_F)}$ |

Comparing the results in Tables 1 and 2, one may notice that while the SNR[$\beta$] and $\mathrm{Q}_{S,3D}^2[\beta]$ in some cells of Table 2 depend on the contrast ($\bar{\mu}$) produced by the feature of interest, SNR[$I$] and $\mathrm{Q}_{S,2D}^2[I]$ in Table 1 depend only on the $\gamma$ value of the feature. This difference in the results is a consequence of the difference in definitions of SNR and $Q_S$ in the 2D and 3D cases. In the case of planar PBI images, it makes theoretical and practical sense to measure the SNR in flat areas of the images, where the variation of pixel intensity values is unaffected by the variation of the refractive index in the imaged object. On the other hand, in the case of CT, each individual 2D projection is usually divided by the corresponding flat-field image (collected without the sample), and then a logarithm of the ratio is taken in the process of CT reconstruction. This means that the mean value of the reconstructed distribution of $\beta$ in flat areas outside the sample is equal to zero (the logarithm of unity). Consequently, the SNR cannot be meaningfully defined in object-free areas. Instead, we calculate the SNR in PB-CT as the ratio of the mean value to the standard deviation of the reconstructed values of $\beta$ in flat areas inside the feature of interest, where the mean reconstructed value is $\bar{\beta} = \bar{\mu}\lambda / (4\pi)$.

## 7. TIE-Hom image reconstruction combined with PSF deconvolution

It is generally accepted that, in the context of phase-contrast imaging and CT, the optimal way to handle the phase contrast is to apply a "phase retrieval" procedure that reconstructs the projected phase distribution from the detected phase-contrast image (Nugent et al., 1996; Paganin, 2006; Wilkins et al., 2014). In the case of PBI and PB-CT of homogeneous samples, such an approach corresponds to application of the TIE-Hom retrieval to the detected propagated image. However, in the presence of stochastic image noise and finite spatial resolution, the situation is less straightforward, and it is necessary to specify what is understood by "optimal solution" first, before attempting to find such a solution. In the published literature, optimal solutions have been previously defined in terms of the maximal SNR (Teague, 1983; Nesterets et al., 2014), the finest spatial resolution (Gureyev et al., 2004a, Beltran et al., 2018) or the maximum ratio of the SNR to the spatial resolution (Gureyev et al., 2014, 2016). Other metrics, such as the previously mentioned "intrinsic imaging quality characteristic" $Q_S$ and the related "biomedical X-ray imaging quality characteristic" were also introduced and shown to be closely related to the Shannon information capacity of the imaging system (Shannon 1948, 1949; Gureyev et al., 2016, 2025). Recent discussions about different types of image quality metrics in biomedical imaging can be found in (Chow et al., 2016; Breger et al., 2025; Gureyev et al., 2026b). In this section, we consider image reconstruction strategies that can potentially produce higher values of $Q_S$ than the conventional TIE-Hom method in PBI and PB-CT.

As shown in Table 1, the $\mathrm{Q}_{S,2D}^2$ value produced by the TIE-Hom reconstruction is equal to $\eta e^{-B_0}(\gamma / N_F)$. On the other hand, the value of $\mathrm{Q}_{S,2D}^2$ after a coherent free-space propagation, before the TIE-Hom reconstruction, is substantially higher: $2\eta e^{-B_0}(\gamma / N_F)^2$. We conjecture that the reduction of the image quality after the TIE-Hom reconstruction is due to a sub-optimal treatment of the PSF of the imaging system in the TIE-Hom reconstruction. This idea has been previously considered and partially addressed in (Gureyev et al., 2004a; Beltran et al., 2018; Alloo et al., 2022)[8]. A modification of the TIE-Hom method that explicitly takes the PSF into account was considered in (Gureyev et al., 2026b). In this method, the following reconstruction operation is applied in place of eq.(8):

$$\bar{I}_{rec,G}(\mathbf{r}_\perp) = [(1 - a^2\nabla_\perp^2)G(\mathbf{r}_\perp, \sigma_{det})*]^{-1}\bar{I}_R(\mathbf{r}_\perp), \tag{25}$$

where the inversion operator on the right-hand side of eq.(25) can be expressed with the help of the Fourier transform: $[(1-a^2\nabla_\perp^2)G(\mathbf{r}_\perp,\sigma_{det})*]^{-1} f(\mathbf{r}_\perp) = \iint \frac{\exp(i2\pi\mathbf{r}_\perp \cdot \boldsymbol{\rho}_\perp)\hat{f}(\boldsymbol{\rho}_\perp)}{(1+4\pi^2 a^2\rho_\perp^2)\hat{G}(\boldsymbol{\rho}_\perp,\sigma_{det})} d\boldsymbol{\rho}_\perp$.

Using eq.(11) with the filter $\hat{T}_{inv,G}(\boldsymbol{\rho}_\perp, a, \sigma_{det}) \equiv (1+4\pi^2 a^2\rho_\perp^2)^{-1}\hat{G}^{-1}(\boldsymbol{\rho}_\perp,\sigma_{det})$ in place of $\hat{T}_{inv}(\boldsymbol{\rho}_\perp, a)$, we obtain the following expression for noise variance in flat areas of the reconstructed image:

$$\mathrm{Var}[I_{rec,G}] = \bar{I}_{tr}\iint \frac{d\boldsymbol{\rho}_\perp}{(1+4\pi^2 a^2\rho_\perp^2)^2} = \frac{\bar{I}_{tr}}{4\pi a^2} = \frac{\bar{I}_{tr}}{\gamma R\lambda} = \frac{\bar{I}_{tr}}{\Delta_{det}^2}\frac{N_F}{\gamma}, \tag{26}$$

which coincides with the first additive term on the right-hand side of eq.(12). The mean signal reconstructed in accordance with eq.(25) is still equal to $\bar{I}_{tr}$, and hence the $\mathrm{SNR}^2$ is equal to

$$\mathrm{SNR}^2[I_{rec,G}] \equiv \bar{I}_{rec,G}^2 / \mathrm{Var}[I_{rec,G}] = \bar{I}_{tr}\gamma R\lambda = \mathrm{SNR}^2[I_0](\gamma / N_F), \tag{27}$$

where we used the fact that $\mathrm{SNR}^2[I_0] = \bar{I}_{tr}\Delta_{det}^2$, according to eq.(13a). Comparing eq.(27) with eq.(13b), we see that, when $\gamma / N_F >> 1$, the gain in SNR in the case of the TIE-Hom reconstruction with PSF deconvolution is slightly smaller than in the case of the original TIE-Hom. This is an expected result, since we have effectively reversed the convolution of the detected signal with the system PSF in the new method. However, the elimination of the convolution with the system PSF leads to a gain in the spatial resolution. Indeed, after the modified TIE-Hom reconstruction, the PSF in the object plane becomes equal to the 2D Dirac delta function: $P_{rec,G}(\mathbf{r}_\perp) = [(1-a^2\nabla_\perp^2)G(\mathbf{r}_\perp,\sigma_{det})*]^{-1}P_R(\mathbf{r}_\perp) =$ $[(1-a^2\nabla_\perp^2)G(\mathbf{r}_\perp,\sigma_{det})*]^{-1}(1-a^2\nabla_\perp^2)G(\mathbf{r}_\perp,\sigma_{det}) = \delta_D(\mathbf{r}_\perp)$. Therefore, at least formally, the spatial resolution in the reconstructed image can become infinitely fine. Note that, in the case of Dirac's delta-function, the spatial resolution cannot be directly calculated using the definition given by eq.(15), because the square of the delta-function is undefined. However, if we approximate the delta-function by a sequence of progressively narrower Gaussians, $G(\mathbf{r}_\perp,\sigma)$ with $\sigma \to 0$, it is easy to see that $\Delta[G(\mathbf{r}_\perp,\sigma)] = 1/\|G(\mathbf{r}_\perp,\sigma)\|_2 = 2\sqrt{\pi}\sigma \to 0$. Consequently, the gain in the intrinsic imaging quality becomes infinitely large after the TIE-Hom reconstruction with PSF deconvolution: $\mathrm{Q}_{S,2D}^2 = \mathrm{SNR}^2[I_{rec,G}]/(\bar{I}_{in}\Delta^2[P_{rec,G}]) = \infty$. However, in reality, achieving infinitely fine spatial resolution in the modified TIE-Hom reconstruction is impossible. Indeed, the TIE-Hom approximation is accurate only for sufficiently low spatial frequencies in the image, as described by eq.(5), and it should not be applied to spatial frequencies above the corresponding upper threshold.

[8] The conceptual picture motivating these papers is as follows. For homogeneous-sample X-ray PBI, free-space propagation sharpens measured intensity distributions (relative to the unpropagated distributions) in accord with eq.(4). Indeed, at this level of approximation, the PBI contrast may be viewed in mathematical terms as Laplacian-type image sharpening (cf. Fig. 3(c) below). Conversely, the TIE-Hom phase-retrieval step in eq.(8) is a low-pass filter ("Paganin filter") that suitably blurs the detected PBI image in order to invert/decode the previously mentioned sharpening. Since a nonzero-width PSF also serves to blur intensity images, it follows that "one needs to blur less" when performing TIE-Hom phase retrieval, since "the imaging-system PSF has already performed some of the blurring required for phase retrieval".

In view of the above, it is appropriate to consider a partial deconvolution of the detector PSF using a narrower Gaussian kernel with the standard deviation $\sigma_\kappa \equiv \kappa\sigma_{det}$, $0 < \kappa < 1$. Then $P_{rec',G}(\mathbf{r}_\perp) \equiv [(1-a^2\nabla_\perp^2)G(\mathbf{r}_\perp,\sigma_\kappa)*]^{-1}\,(1-a^2\nabla_\perp^2)G(\mathbf{r}_\perp,\sigma_{det}) = G(\mathbf{r}_\perp,\sigma_{rec'})$, where $\sigma_{rec'} = (\sigma_{det}^2 - \sigma_\kappa^2)^{1/2} = (1-\kappa^2)^{1/2}\sigma_{det}$, and the spatial resolution in the reconstructed image will be equal to $\Delta[P_{rec',G}] = \Delta_{rec'} = (1-\kappa^2)^{1/2}\Delta_{det}$. This resolution must not be finer than the finest spatial details allowed by the TIE-Hom validity condition eq.(5), $\Delta_{obj}^2 > R\lambda$, which implies an approximate condition $(1-\kappa^2)\Delta_{det}^2 \geq R\lambda$. Similarly to eq.(12), it is possible to show that the variance of noise in the image reconstructed using the modified TIE-Hom method, with the deconvolution kernel $G(\mathbf{r}_\perp,\sigma_\kappa)$, will be equal to

$\mathrm{Var}[I_{rec',G}] \cong (\bar{I}_{tr}/\Delta_{rec'}^2)\{(N_{F,rec'}/\gamma) + (N_{F,rec'}/\gamma)^2[\ln(N_{F,rec'}/\gamma) + \gamma_{EM}]$, where $N_{F,rec'} \equiv \Delta_{rec'}^2/(\lambda R) = (1-\kappa^2)N_F$ and we omitted the asymptotically smaller terms of the order of $(N_{F,rec'}/\gamma)^3\ln(N_{F,rec'}/\gamma)$ and higher. Consequently,

$$\mathrm{SNR}^2[I_{rec',G}] \cong \bar{I}_{tr}\Delta_{rec'}^2[(\gamma/N_{F,rec'}) + \ln(\gamma/N_{F,rec'}) - \gamma_{EM}], \tag{28}$$

$$\frac{\mathrm{SNR}^2[I_{rec',G}]}{\Delta^2[P_{rec',G}]} \cong \frac{\bar{I}_{tr}\gamma R\lambda}{(1-\kappa^2)\Delta_{det}^2} \le \bar{I}_{tr}\,\gamma\,, \tag{29}$$

when $\gamma / N_{F,rec'} >> 1$, where we used the condition $(1-\kappa^2)\Delta_{det}^2 \ge R\lambda$ to obtain the last inequality in eq.(29). Choosing the strongest partial deconvolution compatible with the TIE-Hom validity condition, so that $\Delta^2[P_{rec',G}] \cong R\lambda$ and $N_{F,rec'} \cong 1$, we obtain the limiting value:

$$\mathrm{Q}_{S,2D}^2 = \mathrm{SNR}^2[I_{rec',G}] / (\bar{I}_{in}\,\Delta^2[P_{rec',G}]) \cong \eta e^{-B_0}\gamma\,. \tag{30}$$

The value of $\mathrm{Q}_{S,2D}^2$ in eq. (30) is larger by the factor $N_F$ than the corresponding value in the case of conventional TIE-Hom retrieval given in the last cell of Table 1. The gain factor $\mathrm{Q}_{S,2D}[\mathrm{PBI}] / \mathrm{Q}_{S,2D}[\mathrm{ABI}] = \gamma^{1/2}$, which is achieved in eq.(30), depends only on the relative strength of the X-ray refraction and absorption in the imaged feature. Note, however, that we showed earlier in Section 5 that $\mathrm{Q}_{S,2D}^2 \cong 2\eta e^{-B_0}(\gamma / N_F)^2 < 2\eta e^{-B_0}\gamma^2$ in the image plane. The last value is still larger than the value $\mathrm{Q}_{S,2D}^2 = \eta e^{-B_0}\gamma$ in eq.(30). The reasons for this loss of imaging quality at the TIE-Hom retrieval stage are considered next.

## 8. CTF-Hom image reconstruction combined with PSF deconvolution

It was shown in (Cowley, 1995; Pogany et al., 1997; Gureyev et al., 2004b, 2006) that in the case of a homogeneous object that can be represented as a sum of a small (but possibly rapidly varying) and a slowly varying (but possibly large) components, the Fourier transform of the propagated transmitted intensity in the Fresnel region can be described by the following expression:

$$|U|^2(\boldsymbol{\rho}_\perp, R) = [\cos(\pi R\lambda\rho_\perp^2) + \gamma\sin(\pi R\lambda\rho_\perp^2)]|U|^2(\boldsymbol{\rho}_\perp, 0)\,, \tag{31}$$

which is known as the first Born or Contrast Transfer Function approximation in the case of homogeneous objects (CTF-Hom). Note that $\cos(\pi R\lambda\rho_\perp^2) + \gamma\sin(\pi R\lambda\rho_\perp^2) = (1+\gamma^2)^{1/2}\cos(\pi R\lambda\rho_\perp^2 - \arctan\gamma)$. Unlike the TIE-Hom approximation, described by eq.(4) with the validity conditions in eq.(5), the CTF-Hom approximation does not impose any limits on the propagation distance or the minimal size of the imaged feature (the relationship between the two approximations is discussed in the Appendix). The "contrast transfer function", namely the term in square brackets in eq.(31), is sketched as the solid blue line in Fig. 2 for the case where $\gamma = 1000$. Note that, in this figure, the horizontal axis is a dimensionless scaled radial frequency, with the left vertical axis corresponding to the indicated transfer function.

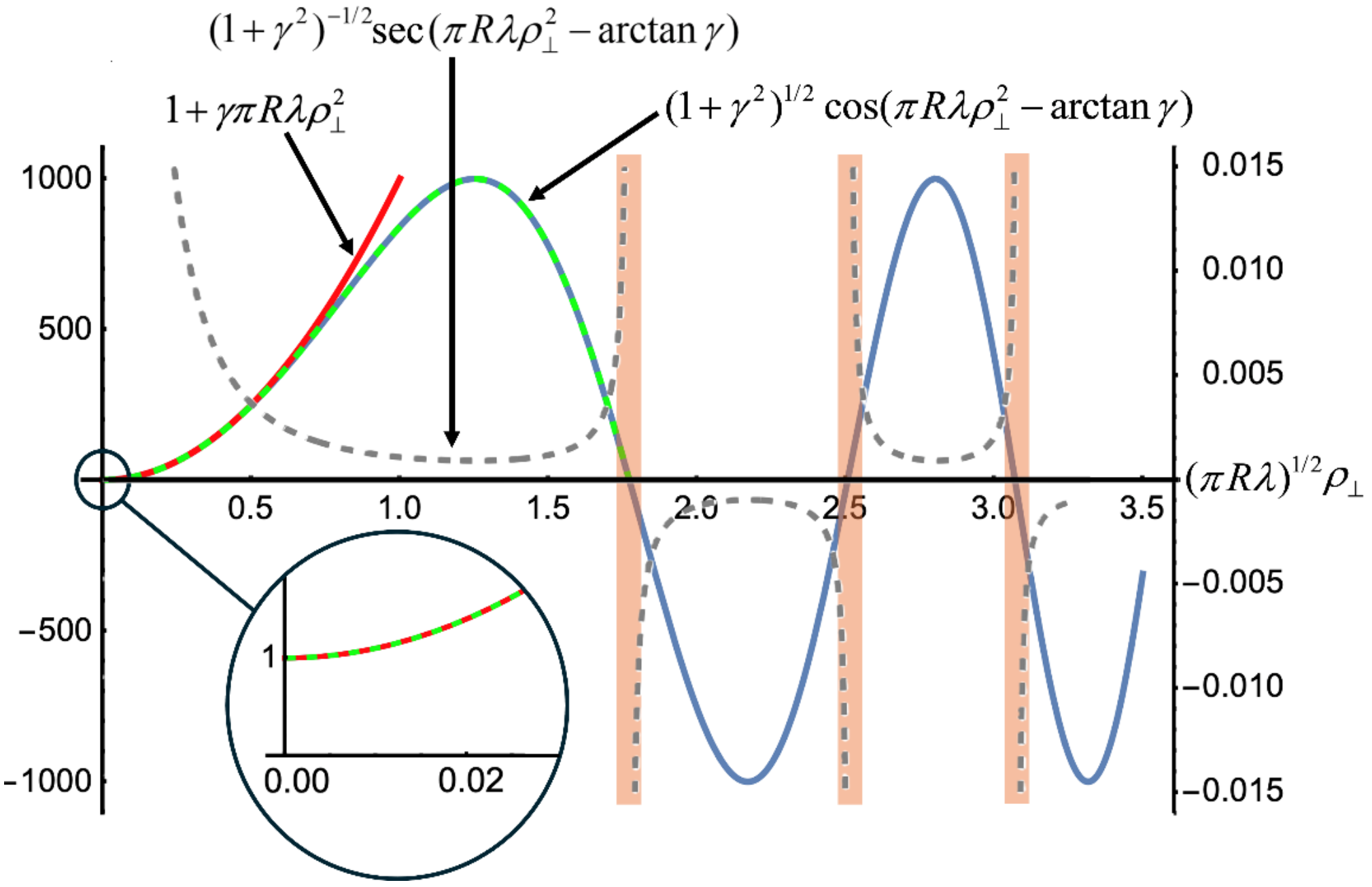

**Fig.2.** Forward and inverse CTF-Hom filter functions, for the case $\gamma = 1000$.

It is easy to see that, in the case of CTF-Hom, the PSF in the image plane is equal to $P_{R,CTF}(\mathbf{r}_\perp) = \{\cos[(4\pi)^{-1}R\lambda\nabla_\perp^2] + \gamma\sin[(4\pi)^{-1}R\lambda\nabla_\perp^2]\}G(\mathbf{r}_\perp, \sigma_{det})$. Let us calculate the corresponding spatial resolution in the image plane using eq.(15). Similarly to TIE-Hom, the description of free-space propagation in terms of the CTF-Hom transformation respects the photon fluence conservation law: $\iint P_{R,CTF}(\mathbf{r}_\perp)d\mathbf{r}_\perp = \hat{P}_{R,CTF}(0) = 1$. Therefore, we only need to calculate the second integral norm of this function in order to evaluate the spatial resolution:

$$\| P_{R,CTF} \|_2^2 = (1+\gamma^2)\iint \cos^2(\pi R\lambda\rho_\perp^2 - \arctan\gamma)\exp(-4\pi^2\sigma_{det}^2\rho_\perp^2)d\boldsymbol{\rho}_\perp =$$

$$\pi(1+\gamma^2)\int_0^\infty \cos^2(\pi R\lambda t - \arctan\gamma)\exp(-4\pi^2\sigma_{det}^2 t)dt = (1+\gamma^2)(R\lambda)^{-1}\exp(-N_F\arctan\gamma)\times$$

$$\int_{-\arctan\gamma}^{\infty} \cos^2(s)\exp(-N_F s)ds = [2(1+\gamma^2) + N_F^2 + 2N_F\gamma]/[\Delta_{det}^2(4+N_F^2)].$$

When $\gamma >> N_F$, the last expression can be accurately approximated by $\| P_{R,CTF} \|_2^2 \cong 2\gamma^2 / [\Delta_{det}^2 (N_F^2 + 4)]$, which implies that

$$\Delta^2[P_{R,CTF}] = \frac{1}{\| P_{R,CTF} \|_2^2} \cong \frac{\Delta^2[P_0](N_F^2+4)}{2\gamma^2}. \quad (32)$$

This spatial resolution asymptotes to $\Delta^2[P_{R,CTF}] = 2\Delta^2[P_0]/\gamma^2$, when $N_F \to 0$, representing the best possible resolution achievable in the image plane in PBI when the propagation distance is no longer restricted to the near-Fresnel region as in the case of the TIE.

In the case of the CTF-Hom approximation, the SNR in the image plane is still equal to that in the object plane, as the photon fluence is conserved and the post-detection PSF is the same in the image plane as in the object plane. Therefore, similar to eq.(14), we have $\mathrm{SNR}^2[I_{R,CTF}] = \mathrm{SNR}^2[I_0] = \bar{I}_{tr}\Delta_{det}^2$ and

$$\frac{SNR^2[I_{R,CTF}]}{\Delta_{R,CTF}^2} \cong \frac{2\bar{I}_{tr}\gamma^2}{N_F^2+4}. \quad (33)$$

The expression in eq.(33) is quadratic in $\gamma$ and reduces to the equivalent formula in the TIE-Hom case, eq.(18a), when $N_F >> 1$. However, unlike the case of eq.(18a), $N_F$ can be arbitrarily small in eq.(33) without that equation diverging. Instead, we have $SNR^2[I_{R,CTF}] / \Delta_{R,CTF}^2 \to \bar{I}_{tr}\gamma^2 / 2$, when $N_F \to 0$, e.g. when the propagation distance tends to infinity, $R \to \infty$. This corresponds to a gain in image quality, in comparison with ABI, that is equal to $\gamma / \sqrt{2}$, which obviously depends only on $\gamma$ and is linear with respect to this parameter.

By analogy with eq.(25), let us consider the PBI image reconstruction performed in accordance with the CTF-Hom approach in combination with the deconvolution of the detector PSF:

$$\hat{I}_{rec}(\boldsymbol{\rho}_\perp) = \hat{I}_R(\boldsymbol{\rho}_\perp) / [(1+\gamma^2)^{1/2} \cos(\pi R\lambda\rho_\perp^2 - \arctan\gamma)\hat{G}(\boldsymbol{\rho}_\perp, \sigma_{det})]. \quad (34)$$

The main difficulty in using this reconstruction method is caused by the presence of zeros of the CTF-Hom in the denominator at points $\rho_\perp$ such that $\cos(\pi R\lambda\rho_\perp^2 - \arctan\gamma) = 0$. Three such zeros can be seen in Fig. 2, at the points where the shaded rectangles cross the horizontal axis. The inverse CTF, shown as the grey dashed lines in Fig. 2, "blows up" at the zeros of the forward CTF (note that the inverse CTF corresponds to the axis on the right side of Fig. 2). This problem can be handled using a regularization method, such as e.g. Tikhonov regularization (Natterer, 2001; Nesterets et al., 2014). This, however, inevitably introduces some distortions in the reconstructed image. The problem with the CTF zeros can be overcome in a more accurate way by collecting PBI images at several different propagation distances $R_n$, which then allows one to effectively eliminate the zeros of the CTF-Hom in the reconstruction process, using the fact that the locations of CTF zeros vary with the propagation distance (Schiske, 2002; Cloetens et al., 1999, Paganin et al., 2004a). However, in the present study we focus on PBI and PB-CT methods that utilize only a single propagation distance. In this case, one approach to avoiding noise amplification at spatial frequencies corresponding to zeros of the CTF-Hom is to limit the range of frequencies to a vicinity of the origin of coordinates in the reciprocal space inside which the CTF-Hom is non-zero. As the first zero of the cosine function occurs at $\pi/2$, the relevant zero-free vicinity can be defined via the inequality $\pi R\lambda\rho_\perp^2 - \arctan\gamma < \pi/2$. This is sketched using the green dotted line in Fig. 2. Therefore, the circle around the point $\boldsymbol{\rho}_\perp = 0$ inside which the CTF-Hom is non-zero has an approximate radius $\rho_{\perp,\max,\varepsilon}^2 = [1/2 + (\arctan\gamma - \varepsilon)/\pi]/(R\lambda)$, where $0 < \varepsilon << 1$ is a small free parameter. Note that, when $\gamma$ is large, $\arctan\gamma$ is close to $\pi/2$. For example, for $\gamma = 1000$, which is a typical order-of-magnitude value in hard X-ray PBI of soft biological tissues, we have $|\pi/2 - \arctan\gamma| < 10^{-3}$ and hence $\rho_{\perp,\max,\varepsilon} \cong 1/(R\lambda)^{1/2}$. If we perform CTF-Hom reconstruction of the propagated image according to eq.(34) only within the frequency range $\rho_\perp < \rho_{\perp,\max,\varepsilon}$, the Fourier transform of the reconstructed PSF, $\hat{P}_{rec,CTF}(\boldsymbol{\rho}_\perp)$, will be equal to unity within the range $\rho_\perp < \rho_{\perp,\max,\varepsilon}$, and it will be equal to the original (uncorrected) PSF in the image plane, $\hat{P}_{R,CTF}(\boldsymbol{\rho}_\perp) = (1+\gamma^2)^{1/2}\cos(\pi R\lambda\rho_\perp^2 - \arctan\gamma)\exp(-2\pi^2\sigma_{det}^2\rho_\perp^2)$, for $\rho_\perp \geq \rho_{\perp,\max,\varepsilon}$. It is easy to see that when $\rho_\perp \geq \rho_{\perp,\max,\varepsilon}$, we have $|\hat{P}_{R,CTF}(\boldsymbol{\rho}_\perp)| \lesssim (1+\gamma^2)^{1/2}\exp(-4N_F/\pi)$. Therefore, when also $N_F > \ln\gamma >> 1$, we have $|\hat{P}_{R,CTF}(\boldsymbol{\rho}_\perp)| << 1$. Therefore, when $N_F > \ln\gamma$, we can approximate $\hat{P}_{rec,CTF}(\boldsymbol{\rho}_\perp)$ by a function which is equal to unity inside the range $\rho_\perp < \rho_{\perp,\max,\varepsilon}$ and is equal to zero outside this range. As $\int P_{rec,CTF}(\mathbf{r}_\perp)d\mathbf{r}_\perp = \hat{P}_{rec,CTF}(0) = 1$ and $\| P_{rec,CTF} \|_2^2 \cong \pi\rho_{\perp,\max,\varepsilon}^2 = (\pi/2 + \arctan\gamma - \varepsilon)/(R\lambda)$, we obtain from eq.(15) that

$$\Delta^2[P_{rec,CTF}] \cong R\lambda/(\pi/2 + \arctan\gamma - \varepsilon) \cong R\lambda/\pi. \quad (35)$$

We see that, according to eq.(35), this "truncated" CTF-Hom retrieval allows one to correctly reconstruct smaller object features compared to the TIE-Hom method, where the minimal feature size was limited by eq.(5). This is an expected outcome, since the TIE-Hom regime corresponds to the small "linear" region of the CTF-Hom near the origin of coordinates, in which $\pi R\lambda\rho_\perp^2 << 1$ and hence $\cos(\pi R\lambda\rho_\perp^2) + \gamma\sin(\pi R\lambda\rho_\perp^2) \cong 1 + \gamma\pi R\lambda\rho_\perp^2$ (see Appendix and the red line in Fig.2). The non-zero region of the CTF-Hom is wider, extending almost up to the boundary of the circle $R\lambda\rho_\perp^2 = 1$.

Let us estimate the variance of noise in the images reconstructed using the CTF-Hom method, eq.(34), when the spatial frequency bandwidth is limited to $\rho_\perp < \rho_{\perp,\max,\varepsilon}$. With reference to eq.(11), we can obtain:

$$\mathrm{Var}[I_{rec,CTF}] = \bar{I}_{tr}(1+\gamma^2)^{-1} \iint_{\rho_\perp < \rho_{\perp,\max,\varepsilon}} \sec^2(\pi R\lambda\rho_\perp^2 - \arctan\gamma)d\boldsymbol{\rho}_\perp =$$

$$2\pi\bar{I}_{tr}(1+\gamma^2)^{-1} \int_0^{\rho_{\perp,\max,\varepsilon}} \sec^2(\pi R\lambda u^2 - \arctan\gamma)u du = \bar{I}_{tr}[(1+\gamma^2)(R\lambda)]^{-1} \int_{-\arctan\gamma}^{\pi/2-\varepsilon} \sec^2(t)dt =$$

$\bar{I}_{tr}[\cot\varepsilon + \gamma]/[(1+\gamma^2)(R\lambda)]$. We see that when $\varepsilon \to 0$, $\cot\varepsilon \to \infty$, and hence $\mathrm{Var}[I_{rec,CTF}] \to \infty$, i.e. the noise can be arbitrarily strongly amplified. This amplification is obviously caused by the CTF reconstruction kernel $(1+\gamma^2)^{-1/2}\cos^{-1}(\pi R\lambda\rho_\perp^2 - \arctan\gamma)$ becoming very large near the "critical" frequency $[(1/2 + \arctan\gamma/\pi)/(R\lambda)]^{1/2}$. Note that the reconstruction filter function must be large at these frequencies, because the forward CTF-Hom function $(1+\gamma^2)^{1/2}\cos(\pi R\lambda\rho_\perp^2 - \arctan\gamma)$ is very close to zero there. As the corresponding spatial resolution is bounded from below by eq.(35), the ratio of $\mathrm{SNR}[I_{rec,CTF}]/\Delta[P_{rec,CTF}]$ becomes very small when the spatial resolution approaches the limit shown in eq.(35).

Let us consider the problem of maximization of the ratio

$$\frac{SNR^2[I_{rec,CTF}]}{\Delta^2_{rec,CTF}} \cong \frac{\bar{I}_{tr}(1+\gamma^2)(\pi/2+\arctan\gamma-\varepsilon)}{\gamma+\cot\varepsilon} \cong \frac{\bar{I}_{tr}(1+\gamma^2)(\pi-\varepsilon)}{\gamma+\cot\varepsilon} \tag{36}$$

as a function of the small positive "regularization" parameter $\varepsilon$. Approximating $\cot\varepsilon \sim 1/\varepsilon$ for small $\varepsilon$, we find that the right-hand side of eq.(36) reaches its maximum at $\varepsilon = (\pi/\gamma)^{1/2}$. When $\gamma >> 1$, $\varepsilon = (\pi/\gamma)^{1/2} << 1$ and the corresponding maximum value of the right-hand side of eq.(36) is approximately equal to

$$\frac{SNR^2[I_{rec,CTF}]}{\Delta^2_{rec,CTF}} \cong \frac{\pi\bar{I}_{tr}(1+\gamma^2)}{\gamma} \cong \pi\bar{I}_{tr}\gamma \text{, and hence } Q^2_{S,2D} \cong \pi\eta e^{-B_0}\gamma . \tag{37}$$

The squared image quality value in eq.(37) is π times larger than the maximum value achievable in the TIE-Hom reconstruction with PSF deconvolution – see eq.(30). Equation (37) indicates that when the imaging setup is not constrained in terms of the allowed propagation distances (or, equivalently, in terms of the spatial frequency range) by the validity conditions of the TIE-Hom approximation, it is possible to overcome the limit previously achieved in eq.(30).

It remains to be checked if it is possible to preserve the image quality gain of the same order as in eq.(33) after a CTF-Hom reconstruction with PSF deconvolution of an image collected at an arbitrary large propagation distance. Note that we already showed that when the range of spatial frequencies was "truncated" in the CTF-Hom reconstruction, and the spatial resolution was limited according to eq.(35), it led to a weaker result for the image quality gain in eq.(37), with the gain factor proportional to $\gamma^{1/2}$. The truncation of the frequency range was applied in order to avoid zeros of the CTF-Hom in the denominator of the image reconstruction filter in eq.(34). These zeros occur at spatial frequencies $\rho_{\perp,m}$, such that $\pi R\lambda\rho^2_{\perp,m} - \arctan\gamma = \pi/2 + m\pi$, where $m = 0,1,2,...$. Instead of truncating the frequency range just before the first zero ($m = 0$), as was done above, we can exclude an $\varepsilon$-vicinity of each "critical frequency" $\rho_{\perp,m}$ without restricting the upper limit of the total frequency range. This corresponds to setting the inverse CTF to zero within the shaded rectangles shown in

Fig. 2. The corresponding distortions of the reconstructed image due to missing spatial frequencies will be reflected in the diminished spatial resolution, as shown below. This procedure of excluding zeros of the CTF-Hom reconstruction filter is similar in effect to the Tikhonov regularization (Natterer, 2001; Nesterets, 2014), but appears better suited for analytical evaluations as we shall demonstrate now.

Consider a similar expression for the noise variance considered above, $\mathrm{Var}[I_{rec,CTF}] = \bar{I}_{tr}(1+\gamma^2)^{-1} \iint_{\Omega_{m_\max,\varepsilon}} \sec^2(\pi R\lambda\rho_\perp^2 - \arctan\gamma) d\boldsymbol{\rho}_\perp$, but this time set the upper limit of spatial frequencies to $\rho_{\perp,m_\max} = (m_{\max} + 1/2 + \arctan\gamma/\pi)^{1/2}/(R\lambda)^{1/2}$, where $m_{\max}$ is a large positive integer number, and exclude the narrow rings $\Omega_{m,\varepsilon} = \left\{ \left( \frac{m+1/2+\arctan\gamma/\pi - \varepsilon/\pi}{R\lambda} \right)^{1/2} < \rho_\perp < \left( \frac{m+1/2+\arctan\gamma/\pi + \varepsilon/\pi}{R\lambda} \right)^{1/2} \right\}$ around each CTF-Hom zero, so that the integration area is $\Omega_{m_\max,\varepsilon} = \{\rho_\perp < \rho_{\perp,m_\max}\} \setminus \bigcup_{m=0}^{m_{\max}} \Omega_{m,\varepsilon}$. As the function under the integral in the expression for $\mathrm{Var}[I_{rec,CTF}]$ is periodic, the whole integral can be estimated via a sum of identical integrals over the individual periods, in addition to the integral over the initial interval containing the point $\rho_\perp = 0$ which was calculated above: $\mathrm{Var}[I_{rec,CTF}] \cong \bar{I}_{tr}[\gamma + (2m_{\max}+1)\cot\varepsilon]/[(1+\gamma^2)(R\lambda)]$. In deriving the last result, we used the same steps as in the above calculation of the noise variance corresponding to CTF-Hom inversion restricted to the spatial frequency interval preceding the first zero of CTF-Hom.

When the highest spatial frequency included in CTF-Hom retrieval is equal to $\rho_{\perp,m_\max,\varepsilon} = (m_{\max} + 1/2 + \arctan\gamma/\pi - \varepsilon/\pi)^{1/2}/(R\lambda)^{1/2}$, the squared spatial resolution after the CTF-Hom retrieval and deconvolution is approximately equal to

$\Delta_{rec,CTF}^2 = \frac{1}{\|\hat{P}_{CTF,rec}\|_2^2} = 1/\iint_{\Omega_{m_\max,\varepsilon}} d\boldsymbol{\rho}_\perp = 1/[\pi\rho_{\perp,m_\max,\varepsilon}^2 - 2\varepsilon m_{\max}/(R\lambda)] = R\lambda/$ $[(2m_{\max}+1)(\pi/2-\varepsilon) + \arctan\gamma]$.

When $m_{\max}$ is allowed to be arbitrarily large, the spatial resolution in the CTF-Hom reconstruction can in principle become as fine as the size of the smallest feature present in the object, $\Delta^2_{obj} = \Delta^2_{rec,CTF} \cong R\lambda / [\pi(m_{\max} + 1)]$, which then determines the ultimate limit of the spatial resolution in the reconstructed image.

Combining the last results for $\mathrm{Var}[I_{rec,CTF}]$ and $\Delta^2_{rec,CTF}$, we obtain:

$$\frac{SNR^2[I_{rec,CTF}]}{\Delta^2_{rec,CTF}} \cong \frac{\bar{I}_{tr}(1+\gamma^2)[\pi(m_{\max}+1) - \varepsilon(2m_{\max}+1)]}{\gamma + (2m_{\max}+1)\cot\varepsilon}. \qquad (38)$$

Note that $m_{\max} + 1$ is associated with the highest order Fresnel fringe detectable in a given PBI setup in the case of an object with extremely sharp edges. When $m_{\max} = 0$, eq.(38) transforms into eq.(36), and, accordingly, the maximum is achieved at $\varepsilon = (\pi / \gamma)^{1/2}$. Consider next an "intermediate" regime, where $1 << m_{\max} << \gamma$. In this case, the right-hand side of eq.(38) can be approximated by $[\bar{I}_{tr}(1+\gamma^2)(\pi - 2\varepsilon)] / [\gamma / m_{\max} + 2\cot\varepsilon]$. This expression reaches its maximum value at $\varepsilon \cong (\pi m_{\max} / \gamma)^{1/2}$, where it becomes

$$\frac{SNR^2[I_{rec,CTF}]}{\Delta^2_{rec,CTF}} \cong \frac{\bar{I}_{tr}\pi m_{\max}(1+\gamma^2)}{\gamma}. \qquad (39)$$

More specifically, in this case we have $SNR^2[I_{rec,CTF}] \cong \bar{I}_{tr} R\lambda(1+\gamma^2)/\gamma$ and $\Delta^2_{rec,CTF} \cong R\lambda / (\pi m_{\max})$. We see that, under such conditions, the spatial resolution in the CTF-Hom reconstructed image can become quite fine, limited from below only by the condition $\Delta^2_{rec,CTF} >> R\lambda / (\pi\gamma)$. The $SNR^2 \sim \bar{I}_{tr} R\lambda\gamma$ remains of the same order with respect of $\gamma$, as in the TIE-Hom case, but can be significantly boosted by the factor $R\lambda / \Delta^2_{det}$, which can be large in the CTF-Hom case.

Finally, for $m_{\max} >> \gamma >> 1$ (e.g. for very long propagation distances), it is possible to show that when $\gamma / m_{\max} \to 0$, the right-hand side of eq.(38) reaches its maximum when $\varepsilon \cong \pi / 2 - [(3/2)\gamma / (2m_{\max}+1)]^{1/3} \to \pi / 2$. The corresponding upper limit is equal to

$$\frac{SNR^2[I_{rec,CTF}]}{\Delta^2_{rec,CTF}} \cong \bar{I}_{tr}\gamma^2 \text{, and hence } Q^2_{S,2D} \cong \eta e^{-B_0}\gamma^2 . \qquad (40)$$

For this choice of the regularization parameter, we get $SNR^2[I_{rec,CTF}] \cong \bar{I}_{tr} R\lambda\gamma^{5/3} / (6m^2_{\max})^{1/3}$ and $\Delta^2_{rec,CTF} \cong R\lambda / (6\gamma m^2_{\max})^{1/3}$, both tending to zero for fixed $\gamma$ when $m_{\max} \to \infty$, but the ratio of $SNR^2$ and $\Delta^2_{rec,CTF}$ staying constant as in eq.(40). Note that, when $\varepsilon \to \pi / 2$, almost all spatial frequencies get excluded from the CTF-Hom reconstruction, except for the frequencies in the narrow vicinity of the lines $\rho_\perp = [(m + \arctan\gamma / \pi) / (R\lambda)]^{1/2}$, where $\cos(\pi R\lambda\rho^2_\perp - \arctan\gamma) \cong 1$. This is a logical outcome of the optimization process which indicates that the highest possible quality of CTF-Hom reconstruction is achieved when imaging samples for which the transmitted intensity in the object plane contains only the spatial frequencies that are transmitted with the highest possible coefficient (close to $(1+\gamma^2)^{1/2}$) to the image plane in the process of forward free-space propagation (see eq.(31)).

**Table 3.** SNR and spatial resolution in 2D distributions of photon fluence in flat areas of images in PBI with and without different reconstruction methods applied, in the case $\gamma$>> 1.

| | $SNR^2[I]$ in flat fields | $\Delta^2$ | $Q^2_{S,2D}$ |
|---|---|---|---|
| Object plane (original) | $\bar{I}_{tr}\Delta^2_{det}$ | $\Delta^2_{det}$ | $\eta e^{-B_0}$ |
| Image plane (TIE-Hom regime) | $\bar{I}_{tr}\Delta^2_{det}$ | $\frac{\Delta^2_{det}}{2(\gamma / N_F)^2}$ | $2\eta e^{-B_0}(\gamma / N_F)^2$ |
| Image plane (CTF-Hom regime) | $\bar{I}_{tr}\Delta^2_{det}$ | $\frac{\Delta^2_{det}(N_F^2 + 4)}{2\gamma^2}$ | $2\eta e^{-B_0}\frac{\gamma^2}{N_F^2 + 4}$ |
| TIE-Hom retrieved | $\bar{I}_{tr}\gamma R\lambda$ | $\Delta^2_{det}$ | $\eta e^{-B_0}(\gamma / N_F)$ |
| TIE-Hom + PSF-Deconv. | $\bar{I}_{tr}\gamma R\lambda$ | $R\lambda$ | $\eta e^{-B_0}\gamma$ |
| CTF-Hom + PSF-Deconv. ($\rho_{\perp,\max} < 1 / \sqrt{R\lambda}$) | $\bar{I}_{tr}\gamma R\lambda$ | $R\lambda / \pi$ | $\pi\eta e^{-B_0}\gamma$ |
| CTF-Hom + PSF-Deconv. ($m_{\max} \cong \rho^2_{\perp,\max} R\lambda >> \gamma >> 1$) | $\frac{\bar{I}_{tr} R\lambda\gamma^{5/3}}{(6m^2_{\max})^{1/3}}$ | $\frac{R\lambda}{(6\gamma m^2_{\max})^{1/3}}$ | $\eta e^{-B_0}\gamma^2$ |

The main results obtained in Sections 5, 7 and 8 are summarized in Table 3. These results indicate, in particular, that the CTF-Hom+PSF-deconvolution approach can deliver superior performance to the TIE-Hom+PSF-deconvolution, in terms of achievable SNR, spatial resolution and the intrinsic imaging quality, $Q_{S,2D}$. These results are verified in the next section using a numerically simulated example.

## 9. Numerical simulations

In this section, we test the theoretical results obtained in Sections 3, 4, 5, 7 and 8 about the SNR and spatial resolution in PBI before and after different variants of phase retrieval. We considered a simple PBI setup (see Fig.1) with a monochromatic plane incident X-ray wave with wavelength $\lambda$ = 1 Å and a sample-to-detector distance $R$ = 10 m. A numerically simulated homogeneous sample had an approximate range of transmission values (0.99, 1.0) and the ratio of the real decrement to the imaginary part of refractive index $\gamma = 100$. The transmitted intensity in the object plane is shown in Fig.3(a), which represents the "ideal" image corresponding to a delta-function PSF and no noise. The detector was assumed to have perfect quantum efficiency, $\eta$ = 1, square pixels with size 1.25 × 1.25 μm$^2$ and a circular Gaussian PSF with $\sigma_{det} = 12.5\ \mu\text{m}$, i.e. with spatial resolution $\Delta_{det} \cong 44.3\ \mu\text{m}$. The registered "flat-field" photon fluence (without the object in the beam) was assumed to be Poisson distributed with a mean value of 1111 photons per detector pixel, which corresponded to noise with the standard deviation of 3 % in the flat-field corrected image. A simulated image intensity in the object plane, corresponding to the stochastic photon fluence convolved with the detector PSF and normalized by the "flat field" image, is shown in Fig.3(b). Figure 3(c) presents the corresponding PBI image in the image plane $z$ = 10 m with the same PSF and photon fluence as in the object plane $z$ = 0. This image was obtained by computing the 2D Fresnel diffraction integrals with the initial complex amplitude corresponding to the intensity distribution in Fig.3(a) and $\gamma = 100$. Comparing the images in Fig.3(b) and Fig.3(c), we see that the image contrast and contrast-to-noise ratio have significantly increased after the coherent free-space propagation.

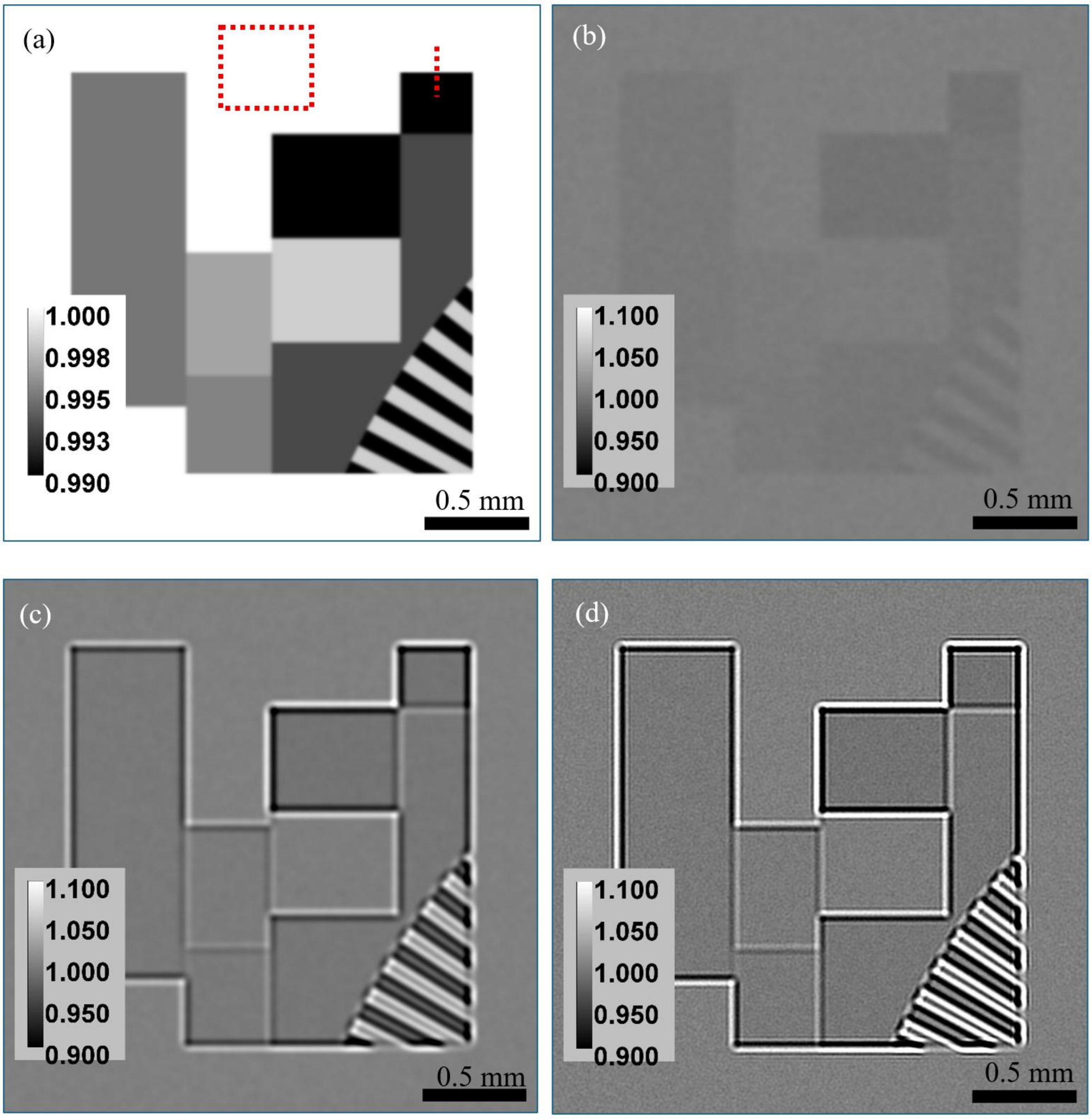

**Fig. 3.** (a) "Ideal" original image in the plane $z = 0$ corresponding to delta-function PSF and no noise. The red dotted line indicates the interval across which the spatial resolution (edge sharpness) was measured in the images. The red dotted rectangle outlines the area inside which the SNR was measured in the images (as the ratio of the average to the standard deviation of pixel values in this area). (b) Detected image in the plane $z = 0$ with 3% Poisson noise and a Gaussian PSF with 12.5 μm standard deviation. (c) Detected image in the plane $z = 10$ m with 3% Poisson noise and a Gaussian PSF with 12.5 μm standard deviation. (d) Same as in (c), but after deconvolution with the regularization parameter $\alpha = 10^{-6}$. The grey scale calibration in the three "detected" images, (b)-(d), is set to the same minimum and maximum levels in order to make the image contrasts easy to compare.

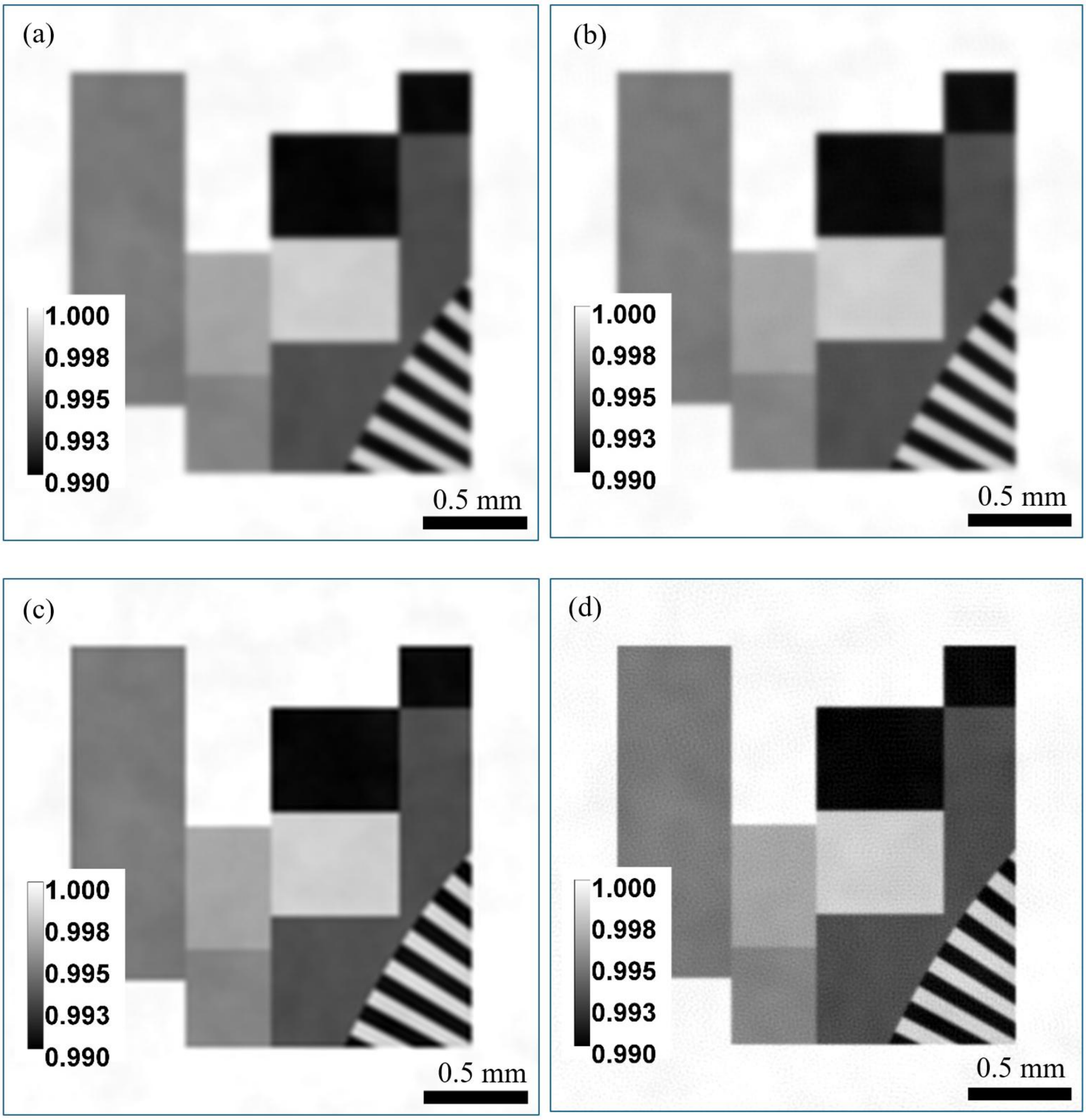

**Fig. 4.** (a) TIE-Hom reconstruction from the propagated image shown in Fig.3(c). (b) CTF-Hom reconstruction from the propagated image shown in Fig.3(c). (c) TIE-Hom reconstruction from the deconvolved propagated image shown in Fig.3(d). (d) CTF-Hom reconstruction from the deconvolved propagated image shown in Fig.3(d). The grey scale calibration in all the images, (a)-(d), was set to the same minimum and maximum levels in order to make the differences in the spatial resolution easier to identify.

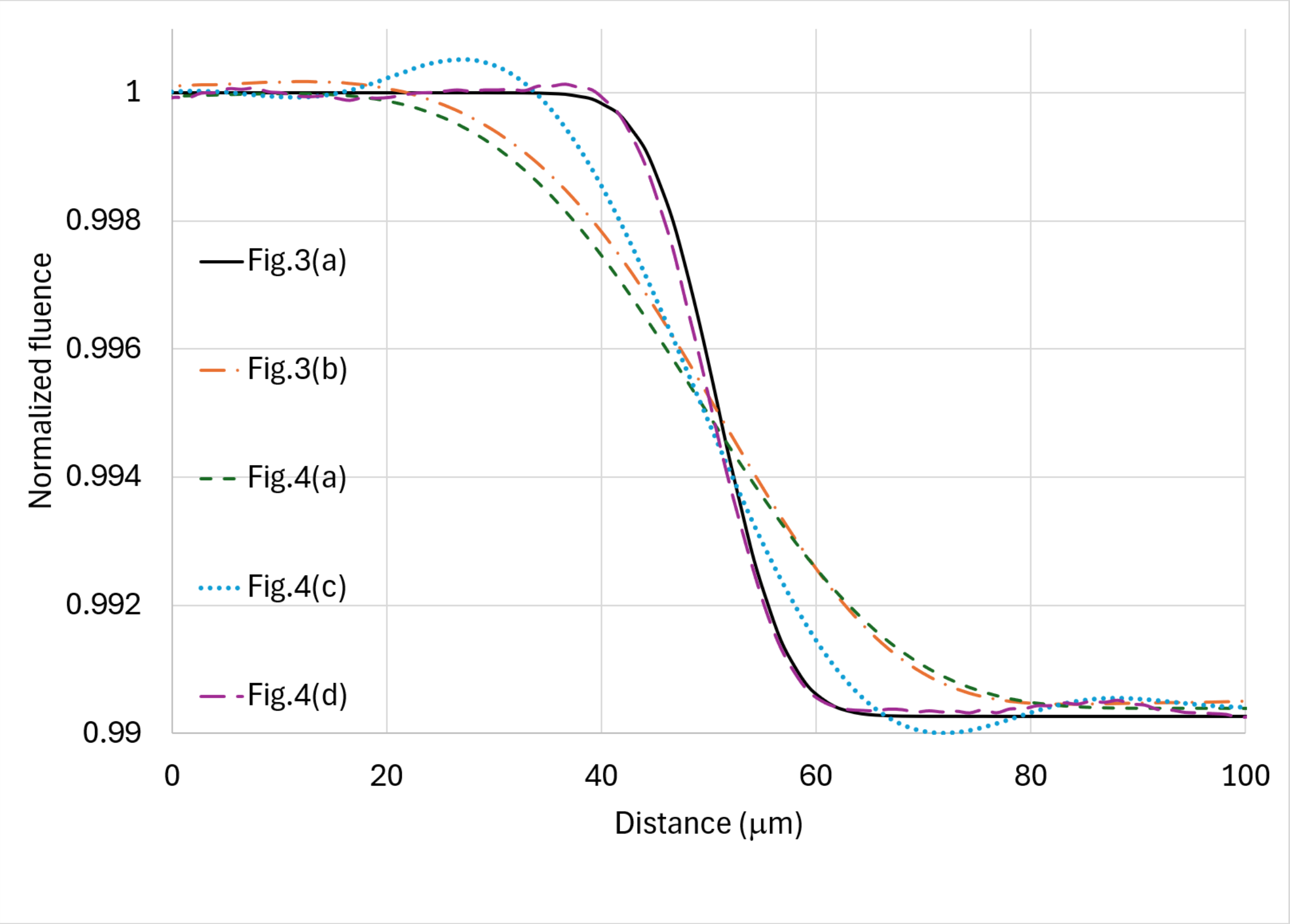

**Fig.5.** Vertical cross-sections along the red dotted line in Fig.3(a), averaged along the horizontal direction, for the images from Fig.3(a) - "ground truth" image in the object plane, (solid black line), Fig.3(b) - detected image in the object plane with 3% Poisson noise and a Gaussian PSF with 12.5 μm standard deviation (orange dot-dash line), Fig.4(a) - TIE-Hom reconstruction from the detected image in the image plane $z$ = 10 m with 3% Poisson noise and a Gaussian PSF with 12.5 μm standard deviation (green short-dash line), Fig.4(c) - same as the previous one, but with the PSF deconvolution (blue dot line), and Fig.4(d) - same as the previous one, but with the CTF reconstruction, instead of the TIE-Hom one (purple long-dash line). These profiles reveal the width of the interface in the images between two adjacent areas with different X-ray absorptions, which allows one to estimate the spatial resolution in the images.

**Table 4.** Theoretical and numerical SNR, spatial resolution and intrinsic imaging quality characteristic in the numerically simulated example images shown in Figs.3-4.

| | Theoretical | | | Numerical | | |
|---|---|---|---|---|---|---|
| | Δ (μm) | SNR | $Q_S$ | Δ (μm) | SNR | $Q_S$ |
| $I_{tr}$ | 1.3 | ∞ | ∞ | 1.3 | $2.6\times10^7$ | $7.7\times10^5$ |
| $I_0$ | 44.3 | 1182 | 1.0 | 42.9 | 1166 | 1.0 |
| $I_R$ | 0.6 | 1182 | 72.0 | 17.9 | 1174 | 2.5 |
| $I_{rec}$ TIE-Hom | 44.3 | 8433 | 7.1 | 44.3 | 8310 | 7.0 |
| $I_{rec}$ TIE-Hom + PSF-deconvolution | 31.6 | 8433 | 10.0 | 27.9 | 8031 | 10.8 |
| $I_{rec}$ CTF-Hom + PSF-deconvolution | 17.8 | 8433 | 17.7 | 12.2 | 6879 | 21.1 |

Note that the Fresnel number here was $N_F \equiv \Delta_{det}^2 / (R\lambda) \cong 2.0$ and the gain factor was $G_2 \equiv (\gamma / N_F)^{1/2} \cong 7.1$ in this example. Figure 3(d) contains the result of Tikhonov-regularized deconvolution, with the regularization parameter $\alpha = 10^{-6}$, of the simulated propagated image shown in Fig.3(c). Figure 3(d) demonstrates a significant increase in the image sharpness (i.e. improved spatial resolution and increased edge contrast) after the deconvolution of the PSF of the imaging system.

The results of the simulated phase retrieval are shown in Fig.4. The top row (panes (a)-(b)) contains the images obtained by the TIE-Hom and CTF-Hom retrieval from the propagated image shown in Fig.3(c), while the bottom row contains images obtained using the same methods with the deconvolved propagated image from Fig.3(d). It is possible to notice that the reconstructed images in Figs.4(c-d), obtained using the combinations of the phase retrieval and PSF deconvolution, are visually sharper (have higher spatial resolution) compared to the corresponding images in Figs.4(a-b) obtained without the deconvolution. This fact can be seen more readily in Fig.5 containing the 1D plots along the dotted red line shown in Fig.3(a) in the images obtained using the TIE-Hom and the CTF-Hom methods with and without the PSF deconvolution. In particular, the line profile through the edge of the image in Fig.4(d), obtained using a combination of CTF-Hom retrieval with the deconvolution of the PSF (purple long-dash line in Fig.5), almost perfectly reproduces the original edge profile in the "ground truth" image in Fig.3(a) (solid black line in Fig.5), without any obvious blurring present in the other line profiles.

We also measured the SNR and spatial resolution in the images shown in Figs.1-2, and included the key results in Table 4, alongside the corresponding analytical results obtained using the expressions from the previous sections of the present paper (see Tables 1-3). One can see that the quantitative results in Table 4 are consistent with the qualitative observations discussed above in relation to Figs.3-5. The SNR, spatial resolution and intrinsic image quality values measured in the simulated images are generally also in good agreement with the theoretical predictions. The apparent exception is represented by the spatial resolution of 17.9 μm measured in the propagated image $I_R$ (shown in Fig.3(c)) that is significantly larger than the corresponding theoretical value of 0.6 μm (see row 3 of Table 4). This discrepancy is due to the practical difficulty in accurately measuring the spatial resolution in the propagated images, where the Fresnel fringes near the edges and interfaces "encode" the improvement in the spatial resolution. This "encoded" fine spatial resolution is subsequently traded for the higher SNR at the phase retrieval step (Gureyev et al., 2017), which, in the present numerical example, worked in agreement with the theory as confirmed by the data in rows 4-6 of Table 4. As predicted by the theoretical estimations in Section 8, the best result in terms of the spatial resolution and intrinsic imaging quality was achieved in the reconstruction combining the CTF-Hom retrieval (in its truncated form – see Section 8) with PSF deconvolution. After the reconstruction using this method, the spatial resolution and the $Q_S$ were both improved by approximately a factor of two after the TIE-Hom retrieval with PSF deconvolution. This result is consistent with the theoretical prediction of achievable spatial resolution of $(R\lambda)^{1/2} \cong 31.6\,\mu\text{m}$ and $(R\lambda/\pi)^{1/2} \cong 17.8\,\mu\text{m}$ in the TIE-Hom + Deconvolution and CTF-Hom + Deconvolution methods, respectively (see Tables 3 and 4). The values of $Q_S$ that were obtained in the numerical example are also in agreement with the theoretical predictions. For example, taking into account that in the present numerical example we have $\eta e^{-B_0} = 1$, $\gamma^{1/2} = 10$ and $(\gamma / N_F)^{1/2} \cong 7.1$, one can see that the values in the last column of Table 4 are generally in good agreement with the analytical expressions in the last column of Table 3. An exception is represented by the measured value of $Q_S \cong 2.5$ in the case of the propagated image $I_R$ in row 3 of Table 4, which was affected by the inaccurate measurement of the spatial resolution in the image plane, the reasons for which were discussed above.

## 10. Conclusions

We recently demonstrated (Gureyev et al., 2026b) that the TIE-Hom+PSF-deconvolution method could produce significantly better results than the original TIE-Hom method, when the imaging quality was measured in terms of the least-squares distance between the reconstructed PBI image and the "ground truth" image obtained in the object plane with no noise and a detector with a delta-function PSF. In (Gureyev et al., 2026b) we also expressed some concerns about using $Q_S$ as a measure of imaging quality in PBI, because this metric could supposedly become infinitely large when PSF deconvolution was used. We have clarified this issue in Section 7 of the present paper by noting that, when the validity conditions of the TIE-Hom method are properly taken into account, the PSF deconvolution must not be applied beyond a certain minimal residual spatial resolution, which is approximately equal to $(R\lambda)^{1/2}$, as follows from eq.(5). When applied properly, with reference to the validity conditions of the used approximations, the $Q_S$ characteristic still provides a correct measure of imaging quality that is closely related to the Shannon information capacity of the imaging method. The usefulness of the $Q_S$ characteristic was further confirmed in the present paper by investigating the application of the CTF-Hom correction method in place of the TIE-Hom retrieval. While the CTF-Hom correction generally requires more care in its application, compared to the TIE-Hom retrieval, due to the presence of zeros in the denominator of the CTF-Hom correction filter, it is capable of accurately reconstructing higher spatial frequencies than TIE-Hom. We showed that the $Q_S$ gain factor relative to ABI can be as high as $(\pi\gamma)^{1/2}$ in the truncated version of the CTF-Hom+PSF-deconvolution method, while in the TIE-Hom+PSF-deconvolution the gain factor could not exceed $\gamma^{1/2}$. In both methods, the single determinant of the gain in image quality was the parameter $\gamma^{1/2}$, unlike the case of the original TIE-Hom method where the gain factor was equal to $(\gamma / N_F)^{1/2}$ and thus also depended on other parameters such as the detector resolution, the propagation distance and the X-ray wavelength. This result shows that: (i) the dependence of the image quality gain on the spatial resolution of the detector can be eliminated by means of deconvolution of the detector PSF, and (ii) the ultimate upper limit of the image quality gain in PBI, in comparison with ABI, is determined solely by the relative strengths of the phase (refraction) and absorption scattering channels.

It was previously shown in (Nesterets et al., 2014), using numerical evaluation of equations for the image quality gain in the case of a Tikhonov regularization of the CTF-Hom retrieval, that the image quality gain factor there could be as high as approximately $0.3\gamma$, which can be higher than $(\pi\gamma)^{1/2}$ when $\gamma$ is large. We proved in Section 8 that a gain in image quality of the order of $\gamma$ can indeed be achieved in a regularized version of the CTF-Hom+PSF-deconvolution method, when the propagation distance is sufficiently long. In particular, we showed that the 2D image quality gain factor can approach $\gamma$ - see eq.(40) or the last row in Table 3. Note, however, that even though in theory this case can deliver the highest gain in imaging quality compared to all other considered results, the corresponding choice of the regularization parameter, $\varepsilon \to \pi/2$, is likely to be unsuitable in practice for most types of imaged samples. Indeed, in such a case a very significant fraction of all spatial frequencies is excluded from the CTF-Hom reconstruction (due to the regularization procedure), so the retrieved image is likely to have extremely strong distortions. It may be preferable to choose instead a smaller value of $\varepsilon$ in the CTF-Hom reconstruction, such as e.g. $\varepsilon = (\pi/\gamma)^{1/2}$ which was found to be optimal in the case of the "truncated" version of the CTF-Hom retrieval above. This will result in a smaller intrinsic imaging quality value compared to eq.(40), but can significantly improve the fidelity of image reconstruction. Note also that in X-ray imaging of biomedical samples with $\gamma >> 1$, it would be very unlikely to encounter the situation corresponding to eq.(40), as it requires $m_{\max} >> \gamma$, i.e. thousands of Fresnel fringes have to be detected (after the PSF deconvolution). A more realistic scenario, even if still relatively rare in practice, corresponds to the case $1 << m_{\max} << \gamma$, i.e. to eq.(39). The corresponding gain in imaging quality in this case is proportional to $\gamma^{1/2}$, as in the TIE-Hom+PSF-deconvolution case. However, the spatial resolution in the CTF-Hom+PSF-deconvolution reconstruction can be much finer compared to the TIE-Hom+PSF-deconvolution case, $\Delta^2_{rec,CTF} \cong R\lambda/(\pi m_{\max})$, $1 << m_{\max} << \gamma$. The most practically relevant case for biomedical imaging is represented by the situation where only a single Fresnel fringe is detectable in PBI images. Here the CTF-Hom+PSF-deconvolution image reconstruction can still potentially deliver an advantage, compared to the TIE-Hom+PSF-deconvolution method. As can be seen by comparing the 5th and the 6th rows of Table 3, while the SNR in the reconstructed image remains the same, both the spatial resolution and the intrinsic imaging quality can be better by a factor of $\pi^{1/2}$ in the case of the CTF-Hom+PSF-deconvolution method compared to the TIE-Hom+PSF-deconvolution method. This advantage was confirmed in the numerical simulation in Section 9.

In the future, it may be interesting to investigate the use of $Q_S$ as a loss metric in training of deep learning (AI) models for tasks such as image deconvolution and denoising (Chen et al., 2017; Pakzad et al., 2026; Tajbakhsh et al., 2026). This option may be particularly beneficial in view of the known low sensitivity of some of the traditionally used loss metrics, such as peak signal-to-noise (PSNR), SSIM and least-squares difference, with a ground truth image, to variations in the propagation phase contrast and phase retrieval (Chen et al., 2017; Pakzad et al., 2026; Gureyev et al., 2026b). This low sensitivity is fundamentally related to the nature of propagation contrast that is typically concentrated at the interfaces and boundaries of the imaged objects, i.e. within the image pixel subsets of co-dimensionality one (“hyper-surfaces”). These regions of noticeable propagation contrast represent image subsets with zero Radon measure, unlike any area-based (or volume-based, in the case of PB-CT) image features corresponding, for example, to regions with different X-ray absorption in ABI. The concept of subsets of zero measure is, of course, an idealization corresponding to continuous images (with an infinitely small pixel size). Nevertheless, in a real digital image, these “hyper-surface” subsets of pixels are naturally expected to contribute very little to any integral image quality metric, such as e.g. the least-squares difference between the reconstructed image and the ground truth. On the other hand, the intrinsic imaging quality characteristic $Q_S$ is very sensitive to both the image noise and the sharpness of edges and interfaces, particularly, if edge sharpness is used directly for the evaluation of spatial resolution in the images. Therefore, $Q_S$ can be expected to perform more efficiently as a loss function driving the training of AI models for denoising, deconvolution and other image processing tasks in PBI.

**Acknowledgements**

The authors acknowledge funding from the Australian National Health and Medical Research Council (grant No. APP2011204).

**Appendix. Validity conditions of the TIE-Hom approximation**

Expanding the sine and cosine functions in eq.(31) into the Taylor series at $\boldsymbol{\rho}_\perp = 0$, one can see that the zero- and first-order terms, arising from the approximations $\cos(\pi R\lambda\rho_\perp^2) \sim 1$ and $\gamma\sin(\pi R\lambda\rho_\perp^2) \sim \pi\gamma R\lambda\rho_\perp^2$, respectively, produce the TIE-Hom equation (4). Consider the next-order expansion terms, $-(1/2)(\pi R\lambda)^2\rho_\perp^4$ and $-(1/6)\gamma(\pi R\lambda)^3\rho_\perp^6$. The TIE-Hom equation represents a good approximation to eq.(31) when the higher-order terms are much smaller in magnitude than at least one of the terms included in eq.(4). Correspondingly, we need to consider the conditions which make the following inequalities correct: $(1/2)(\pi R\lambda)^2\rho_\perp^4 << \pi\gamma R\lambda\rho_\perp^2$ and $(1/3)\gamma(\pi R\lambda)^3\rho_\perp^6 << \pi\gamma R\lambda\rho_\perp^2$. These inequalities can be re-written as $\rho_\perp^2 << 2\gamma/(\pi R\lambda)$ and $\rho_\perp^2 << \sqrt{6}/(\pi R\lambda)$. We are primarily interested in the case of biomedical imaging with hard X-rays, where typically $\gamma >> 1$, and hence $2\gamma >> \sqrt{6}$. Therefore, the latter of the two inequalities for the magnitude of $\rho_\perp^2$ constitutes a tighter restriction on the validity range of the TIE-Hom approximation. We can write this as an approximate upper limit on the spatial frequencies present in the object-plane intensity distribution: $\rho_{\perp,max}^2 < 1/(2R\lambda)$. This can be also expressed as a condition on the minimal size of the projected object features, $\Delta_{obj} \equiv 1/(2\rho_{\perp,max}) > R\lambda$. Thus, a sufficient condition for validity of eq.(4) is

$$\rho_{\perp,max}^2 < 1/(2R\lambda) \text{ or } \Delta_{obj}^2 > R\lambda . \qquad \text{(A1)}$$